\documentclass[journal]{IEEEtran}

\usepackage{cite}

\usepackage{graphicx}
\usepackage{xcolor}
\usepackage{tikz}
\usetikzlibrary{arrows,positioning,shapes.geometric,calc,spy,patterns}
\usepackage{pgfplots}
\usepackage{multirow}
\usepackage{upgreek}
\usepackage{amssymb}
\usepackage{amsmath}
\usepackage{cases}
\usepackage{pifont}
\usepackage{bm}
\usepackage{amsthm}

\usepackage{tabularx}
\usepackage{booktabs}
\usepackage{filecontents}
\usepackage{subcaption}
\usepackage[normalem]{ulem}
\newcolumntype{Y}{>{\centering\arraybackslash}X}

\usepackage[]{algorithm2e}
\usepackage[hyphens]{url}
\SetAlCapSkip{1em}

\usepackage{array}

\DeclareMathOperator*{\sgn}{sgn}

\newcommand{\fixme}[2]{\ifx&#2&{\leavevmode\color{red}#1}\else{\leavevmode\color{red}FIXME\{}#1{\leavevmode\color{red}\}}\footnote{{\leavevmode\color{red}#2}}\PackageWarning{Fixme}{#1: #2}\fi}

\begin{document}
\title{Improved Bit-Flipping Algorithm for Successive Cancellation Decoding of Polar Codes}

\author{Furkan~Ercan,~\IEEEmembership{Student~Member,~IEEE,}
        Carlo~Condo,~\IEEEmembership{Member,~IEEE,}
        and Warren~J.~Gross~\IEEEmembership{Senior~Member,~IEEE}%
\thanks{F.~Ercan and W.~J.~Gross are with the Department of Electrical and Computer Engineering, McGill University, Montr\'eal, Qu\'ebec, Canada. e-mail: furkan.ercan@mail.mcgill.ca, warren.gross@mcgill.ca. C.~Condo was with the Department of Electrical and Computer Engineering, McGill University. Now he is with the Huawei Paris Research Center, Boulogne-Billancourt, France. e-mail: carlo.condo@huawei.com}}

%
%

\markboth{IEEE Transactions on Communications,~Vol.~0, No.~0, August~2015}%
{Shell \MakeLowercase{\textit{et al.}}: Bare Demo of IEEEtran.cls for IEEE Communications Society Journals}
%



\maketitle

\begin{abstract}
The interest in polar codes has been increasing significantly since their adoption for use in the 5$^{\rm th}$ generation wireless systems standard. Successive cancellation (SC) decoding algorithm has low implementation complexity, but yields mediocre error-correction performance at the code lengths of interest. SC-Flip algorithm improves the error-correction performance of SC by identifying possibly erroneous decisions made by SC and re-iterates after flipping one bit. It was recently shown that only a portion of bit-channels are most likely to be in error. In this work, we investigate the average log-likelihood ratio (LLR) values and their distribution related to the erroneous bit-channels, and develop the Thresholded SC-Flip (TSCF) decoding algorithm. We also replace the LLR selection and sorting of SC-Flip with a comparator to reduce the implementation complexity. Simulation results demonstrate that for practical code lengths and a wide range of rates, TSCF shows negligible loss compared to the error-correction performance obtained when all single-errors are corrected. At matching maximum iterations, TSCF has an error-correction performance gain of up to $0.45$ dB compared with SC-Flip decoding. At matching error-correction performance, the computational complexity of TSCF is reduced by up to $40\%$ on average, and requires up to $5\times$ lower maximum number of iterations.

\end{abstract}


\IEEEpeerreviewmaketitle

\section{Introduction}
\IEEEPARstart{P}{olar} codes, introduced by Ar{\i}kan in \cite{arikan09}, are a class of linear block codes that provably achieve channel capacity. Due to their low-complexity encoding and decoding, polar codes have been selected as a coding scheme in the $5^{\rm th}$ generation wireless systems standards (5G). Current use of polar codes within 5G includes the enhanced mobile broadband (eMBB) control channel \cite{3GPP-5G}, while they are being considered for ultra-reliable low-latency communications (URLLC) and massive machine-type communications (mMTC). 5G communications require decoding algorithms that have improved error-correction performance and throughput, along with reduced power and energy consumption compared to previous communication standards.

Successive-cancellation (SC) decoding is the first decoding algorithm for polar codes, proposed in \cite{arikan09}. SC decoding is able to achieve channel capacity as the codeword length tends to infinity; however, its error-correction performance is degraded at practical lengths. To tackle this issue, SC-List decoding algorithm was proposed in \cite{TalList}. SC-List decoders improve the SC error-correction performance at the cost of longer latency, as well as increased area occupation and power consumption. Several contributions have improved latency, area occupation and power consumption of SC-List algorithm \cite{list-LLR,fastSSCL-conf,fastSSCL-jour,PSCL-GLOBECOM,ercan-allerton}. 

The SC-Flip decoding algorithm takes an alternative approach to the improvement of SC \cite{SCFlip14}. In contrast to the parallel SC decoding approach of SC-List, SC-Flip relies on multiple subsequent decoding attempts in order to identify and correct the wrong decision made by SC, sequentially flipping unreliable bits. It yields an average computational complexity equivalent to that of SC decoding at medium to high signal-to-noise ratio (SNR) values, mainly sacrificing worst case latency. Its error-correction performance is comparable to that of SC-List decoding with a list size of $2$ at practical iteration lengths. Modifications to SC-Flip decoding that were carried out in \cite{SCFlip17-conf,SCFlip17-jour,SCF-WCNC18,SCF-GLOBECOM17,PSCF-ICC18} have shown that the error-correction performance can be greatly improved.

In \cite{SCF-WCNC18}, two methods were proposed to improve SC-Flip decoding, based on the distribution of first wrong bit estimation, that is always caused by channel noise. One method simplifies the operations required by SC-Flip decoding, without degrading the error-correction performance. The second method restricts the search space for bit-flipping, improving the error-correction performance. These techniques are very effective at low code rates, but their gain degrades quickly as the code rate increases. 

The focus of this work is to identify and correct the first channel-induced error as efficiently as possible. Our approach is based on a critical set of bit indices that are more error-susceptible than the rest, similar to \cite{SCF-WCNC18}. We summarize the contributions of this work as follows:
\begin{itemize}
\item The log-likelihood ratio (LLR) values of the critical set are observed, noting that the mean values of LLRs of critical set for successful and unsuccessful decoding differ substantially.
\item Using this information on a critical set, an LLR threshold is applied to further restrict the search space for bit-flipping; with this technique, SC-Flip is able to approach the error-correction performance obtained when all first channel-induced errors are corrected, with a small number of iterations at high, medium and low code rates. We name this new decoding approach as Thresholded SC-Flip (TSCF) algorithm. Simulations show that TSCF has negligible loss compared to the error-correction performance that can be achieved when all single-channel-induced errors are corrected. 
\item In TSCF decoding, we replace the the LLR selection and sorting of the original SC-Flip with a simple threshold comparator to further reduce the implementation complexity.
\item We show that although the observation of the critical set depends on the simulated $E_b/N_0$, a single critical set valid for all SNRs is sufficient to substantially improve the error-correction performance compared to the baseline SC-Flip decoding. 
 
\end{itemize}

The rest of this work is organized as follows: In Section \ref{sec:bg}, encoding and decoding techniques of polar codes are reviewed. In Section \ref{sec:ourtech}, the proposed SC-Flip with thresholded index selection is described. Simulation results are presented in Section \ref{sec:simresults}, and conclusions are drawn in Section \ref{sec:conc}.

\section{Preliminaries}\label{sec:bg}

\subsection{Polar Codes}

A polar code $PC(N,K)$ of code length $N$ and rate $R=K/N$ is a linear block code that divides $N = 2^n, n \in \mathbb Z^+$ bit-channels in $K$ reliable ones and $N-K$ unreliable ones. Information bits are transmitted via the reliable channels. The unreliable channels are fixed to a value which is known by both the transmitter and the receiver, usually zero. They are thus called frozen channels.

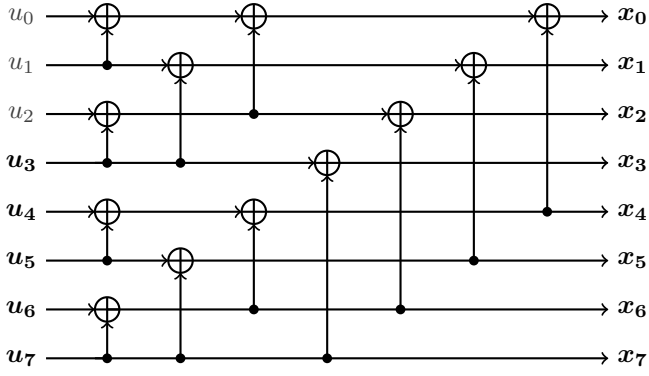
\begin{figure}
   \centering
   \begin{tikzpicture}[scale=.65, thick]
  \node [color=darkgray] at (.5,0) {$u_0$} ;
  \node [color=darkgray]at (.5,-1) {$u_1$};
  \node [color=darkgray]at (.5,-2) {$u_2$};
  \node at (.5,-3) {$\boldsymbol{u_3}$};
  \node at (.5,-4) {$\boldsymbol{u_4}$};
  \node at (.5,-5) {$\boldsymbol{u_5}$};
  \node at (.5,-6) {$\boldsymbol{u_6}$};
  \node at (.5,-7) {$\boldsymbol{u_7}$};

  \foreach \x in {-6,-4,-2,0}
  {
    \draw [->] (1,\x) -- (2,\x);
    \draw (1,\x-1) -- (2.25,\x-1);

    \draw (2.25,\x) circle [radius=.25];
    \draw (2,\x) -- (2.5,\x);
    \draw (2.25,\x-.25) -- (2.25,\x+.25);

    \draw [->] (2.25,\x-1) -- (2.25,\x-.25);

    \fill (2.25,\x-1) circle [radius=.1];
  }

  \foreach \x in {-4,0}
  {
    \draw [->] (2.5,\x) -- (5,\x);
    \draw [->] (2.25,\x-1) -- (3.5,\x-1);

    \draw (5.25,\x) circle [radius=.25];
    \draw (5,\x) -- (5.5,\x);
    \draw (5.25,\x-.25) -- (5.25,\x+.25);

    \draw (3.75,\x-1) circle [radius=.25];
    \draw (3.5,\x-1) -- (4,\x-1);
    \draw (3.75,\x-1-.25) -- (3.75,\x-1+.25);

    \draw [->] (2.25,\x-2) -- (5.25,\x-2) -- (5.25,\x-.25);
    \fill (5.25,\x-2) circle [radius=.1];
    \draw [->] (2,\x-3) -- (3.75,\x-3) -- (3.75,\x-1-.25);
    \fill (3.75,\x-3) circle [radius=.1];
  }

  \draw [->] (5.5,0) -- (11,0);
  \draw [->] (4,-1) -- (9.5,-1);
  \draw [->] (5.25,-2) -- (8,-2);
  \draw [->] (3.75,-3) -- (6.5,-3);

  \foreach \x in {-1,0}
  {
    \draw [->] (5.5+1.5*\x,\x-4) -- (11.25+1.5*\x,\x-4) -- (11.25+1.5*\x,\x-.25);
    \draw [->] (5.25+1.5*\x,\x-6) -- (11.25+1.5*\x-3,\x-6) -- (11.25+1.5*\x-3,\x-2-.25);
  }

  \foreach \x in {-3,...,0}
  {
    \draw (11.25+1.5*\x,\x) circle [radius=.25];
    \draw (11+1.5*\x,\x) -- (11.5+1.5*\x,\x);
    \draw (11.25+1.5*\x,\x-.25) -- (11.25+1.5*\x,\x+.25);

    \fill (11.25+1.5*\x,\x-4) circle [radius=.1];

    \draw [->] (11.5+1.5*\x,\x) -- (12.5,\x);
    \draw [->] (11.25+1.5*\x,\x-4) -- (12.5,\x-4);
  }

  \node at (13,0) {$\boldsymbol{x_0}$};
  \node at (13,-1) {$\boldsymbol{x_1}$};
  \node at (13,-2) {$\boldsymbol{x_2}$};
  \node at (13,-3) {$\boldsymbol{x_3}$};
  \node at (13,-4) {$\boldsymbol{x_4}$};
  \node at (13,-5) {$\boldsymbol{x_5}$};
  \node at (13,-6) {$\boldsymbol{x_6}$};
  \node at (13,-7) {$\boldsymbol{x_7}$};

\end{tikzpicture}
   \caption{Polar code encoding for $PC(8,5)$.}
   \label{fig:polarencode}
 \end{figure}

Polar codes are encoded through the following matrix multiplication:
\begin{equation}\label{eqn:enc}
\boldsymbol{x_0^{N-1}} = \boldsymbol{u_0^{N-1}}G^{\otimes n}\text{,}
\end{equation}
where $\boldsymbol{x_0^{N-1}} = \{x_0,x_1,\ldots,x_{N-1}\}$ represents the encoded vector, $\boldsymbol{u_0^{N-1}} = \{u_0,u_1,\ldots,u_{N-1}\}$ is the input vector, and the generator matrix $G^{\otimes n}$ is obtained as the $n$-th Kronecker product of the polarization matrix $G = \left[\begin{smallmatrix} 1&0\\ 1&1 \end{smallmatrix} \right]$. Due to the recursive nature of the encoding process, an $N$-length polar code can be interpreted as the concatenation of two polar codes of length $N/2$. 
 The encoding operation in (\ref{eqn:enc}) for polar code $PC(8,5)$ is portrayed in Fig. \ref{fig:polarencode}; gray indices represent the frozen bits whereas the black indices indicate the information bits.
 
\begin{figure}
  \centering
   \scalebox{0.75}{\begin{tikzpicture}[scale=.65, thick]

\draw [thin,gray,dashed] (0,-1) -- (7.5,-1);
\draw [thin,gray,dashed] (0,-3) -- (11.5,-3);
\draw [thin,gray,dashed] (0,-5) -- (13.5,-5);
\draw [thin,gray,dashed] (0,-7) -- (14.5,-7);

\node at (-.75,-1) {$S=3$};
\node at (-.75,-3) {$S=2$};
\node at (-.75,-5) {$S=1$};
\node at (-.75,-7) {$S=0$};

\draw (7.5,-1) -- (3.5,-3);
\draw (7.5,-1) -- (11.5,-3);

\draw (3.5,-3) -- (1.5,-5);
\draw (3.5,-3) -- (5.5,-5);
\draw (11.5,-3) -- (9.5,-5);
\draw (11.5,-3) -- (13.5,-5);

\draw (1.5,-5) -- (0.5,-7);
\draw (1.5,-5) -- (2.5,-7);

\draw (5.5,-5) -- (4.5,-7);
\draw (5.5,-5) -- (6.5,-7);

\draw (9.5,-5) -- (8.5,-7);
\draw (9.5,-5) -- (10.5,-7);

\draw (13.5,-5) -- (12.5,-7);
\draw (13.5,-5) -- (14.5,-7);


  \draw[black,fill=white] (.5,-7) circle [radius=.25];			
  \fill[color=gray] (1.5,-5) circle [radius=.25];	
  \draw[black,fill=white] (2.5,-7) circle [radius=.25];			
  \fill[color=gray] (3.5,-3) circle [radius=.25];	
  \draw[black,fill=white] (4.5,-7) circle [radius=.25];			
  \fill[color=gray] (5.5,-5) circle [radius=.25];	
  \fill[color=black] (6.5,-7) circle [radius=.25];	

  \fill[color=gray] (7.5,-1) circle [radius=.25];	

  \fill[color=black] (8.5,-7) circle [radius=.25];		
  \fill[color=gray] (9.5,-5) circle [radius=.25];		
  \fill[color=black] (10.5,-7) circle [radius=.25];		
  \fill[color=gray] (11.5,-3) circle [radius=.25];		
  \fill[color=black] (12.5,-7) circle [radius=.25];		
  \fill[color=gray] (13.5,-5) circle [radius=.25];		
  \fill[color=black] (14.5,-7) circle [radius=.25];		

\node at (.5,-7.8) {$\hat{u}_0$};
\node at (2.5,-7.8) {$\hat{u}_1$};
\node at (4.5,-7.8) {$\hat{u}_2$};
\node at (6.5,-7.8) {$\hat{u}_3$};
\node at (8.5,-7.8) {$\hat{u}_4$};
\node at (10.5,-7.8) {$\hat{u}_5$};
\node at (12.5,-7.8) {$\hat{u}_6$};
\node at (14.5,-7.8) {$\hat{u}_7$};

\draw [->] (8,-1.125) -- (11,-2.625) node [above=.05cm,midway,rotate=-25] {$\boldsymbol{\alpha}$};
\draw [<-] (8,-1.375) -- (11,-2.875) node [below=-.05cm,midway,rotate=-25] {$\boldsymbol{\beta}$};

\draw [->] (11.25,-3) -- (9.75,-4.5) node [above=.03cm,midway,rotate=40] {$\boldsymbol{{\alpha}^l}$};
\draw [<-] (11.25,-3.5) -- (9.75,-5) node [below=-.05cm,midway,rotate=40] {$\boldsymbol{{\beta}^l}$};

\draw [->] (11.75,-3) -- (13.25,-4.5) node [above=.03cm,midway,rotate=-40] {$\boldsymbol{{\alpha}^r}$};
\draw [<-] (11.75,-3.5) -- (13.25,-5) node [below=-.05cm,midway,rotate=-40] {$\boldsymbol{{\beta}^r}$};

\end{tikzpicture}}
  \\
  \vspace{2pt}
  \caption{Successive-cancellation decoding tree for a $PC(8,5)$ code.}
  \label{fig:scdecode}
\end{figure}
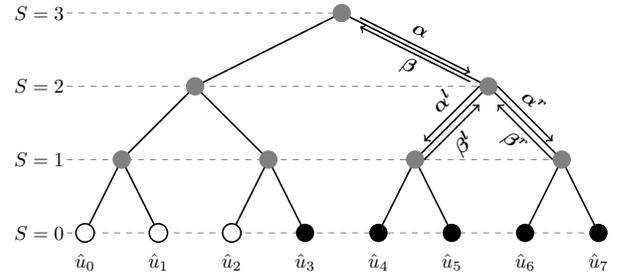

The scheduling of operations required by the SC decoding algorithm allows to see its process as a binary tree search, where the tree is explored depth-first, with priority given to the left branch. 
Fig. \ref{fig:scdecode} portrays an example of SC decoding tree, for $PC(8,5)$. Each node receives from its parent a vector of LLRs $\boldsymbol{\alpha}=\{\alpha_0, \alpha_1, \ldots, \alpha_{2^S-1}\}$. Each node at stage $S$ computes the left $\boldsymbol{\alpha^l} = \{\alpha^l_0, \alpha^l_1, \ldots, \alpha^l_{2^{S-1}-1}\}$ and right $\boldsymbol{\alpha^r} = \{\alpha^r_0, \alpha^r_1, \ldots, \alpha^r_{2^{S-1}-1}\}$ LLR vectors sent to child nodes as
\begin{align}
{\alpha}^l_i &= \sgn(\alpha_{i})\sgn(\alpha_{i+2^{S-1}}) \min(\alpha_{i},\alpha_{i+2^{S-1}}) \text{,} \label{eqn:alphaleft}\\
{\alpha}^r_i &= \alpha_{i+2^{S-1}} + (1-2\beta^{l}_{i})\alpha_{i} \text{.} \label{eqn:alpharight}
\end{align}
The LLRs at the root node are initialized as the channel LLR $\boldsymbol{y}_0^{N-1} = \{y_0,y_1,\ldots,y_{N-1}\}$. Nodes receive the partial sums $\boldsymbol{\beta}$ from their left $\boldsymbol{\beta^{l}}=\{\beta^l_0, \beta^l_1, \ldots, \beta^l_{2^{S-1}-1}\}$ and right $\boldsymbol{\beta^{r}}=\{\beta^r_0, \beta^r_1, \ldots, \beta^r_{2^{S-1}-1}\}$ child node:
\begin{equation}\label{eqn:beta}
  \beta_i=\left\{
  \begin{array}{@{}ll@{}}
    \beta^{l}_{i} \oplus \beta^{r}_{i}, & \text{if}~ i \leq 2^{S-1} \\
    \beta^{r}_{i-2^{S-1}}, & \text{otherwise.}
  \end{array}\right.
\end{equation} 
where $\boldsymbol{\oplus}$ is the bitwise XOR operation, and $0 \leq i < 2^S$. At leaf nodes, the $\beta$ value and the estimated bit vector $\boldsymbol{\hat{u}_0^{N-1}}$ are computed as
\begin{equation}\label{eqn:bitestimate-sc}
\beta_{i}=\left\{
  \begin{array}{@{}ll@{}}
    0, & \text{when } \alpha_i \geq 0 \text{ } \text{or } i \in \Phi; \\
    1, & \text{otherwise.}
  \end{array}\right.
\end{equation}
where $\Phi$ denotes the set of frozen indices.

\subsection{Successive-Cancellation Flip Decoding}\label{sec:SC-Flip}

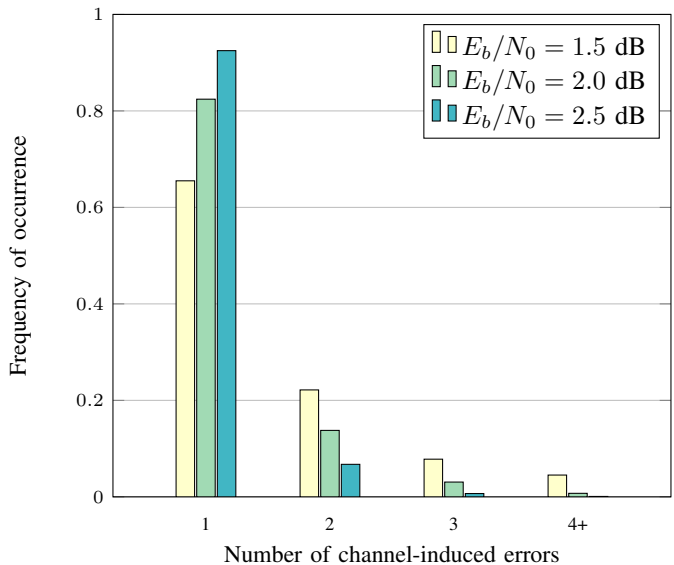
\begin{figure}
   \centering
   \begin{tikzpicture}

 \definecolor{r1}{RGB}{255,255,204}
 \definecolor{r2}{RGB}{161,218,180}
 \definecolor{r3}{RGB}{65,182,196}
 \definecolor{r4}{RGB}{44,127,184}
 \definecolor{r5}{RGB}{37,52,148}

  \pgfplotsset{
    label style = {font=\fontsize{9pt}{7.2}\selectfont},
    tick label style = {font=\fontsize{7pt}{7.2}\selectfont}
  }

\begin{axis}[
	width  = 9cm,
        height = 8cm,
        major x tick style = transparent,
        ybar=2*\pgflinewidth,
        bar width=7pt,
        ymajorgrids = true,
        xlabel = {Number of channel-induced errors},
        ylabel = {Frequency of occurrence},
        symbolic x coords={1,2,3,4+},
        xtick = data,
        scaled y ticks = false,
        enlarge x limits=0.25,
        ymin=0,
        ymax=1,
]

	\addplot[style={black,fill=r1,mark=none}]
            coordinates {(1,0.6551) (2,0.221657) (3,0.0780736) (4+,0.045169)};
            \addlegendentry{$E_b/N_0=1.5$ dB}

        \addplot[style={black,fill=r2,mark=none}]
              coordinates {(1,0.824311) (2,0.13775) (3,0.0306435) (4+,0.00729607)};
             \addlegendentry{$E_b/N_0=2.0$ dB}

        \addplot[style={black,fill=r3,mark=none}]
              coordinates {(1,0.924916) (2,0.067415) (3,0.00682276) (4+,0.000846496)};
             \addlegendentry{$E_b/N_0=2.5$ dB}

\end{axis}
\end{tikzpicture}
   \caption{Frequency of occurrence of channel-induced errors at various $E_b/N_0$ points for $PC(512,256)$.}
   \label{fig:errorlikelihoods-snr}
 \end{figure}

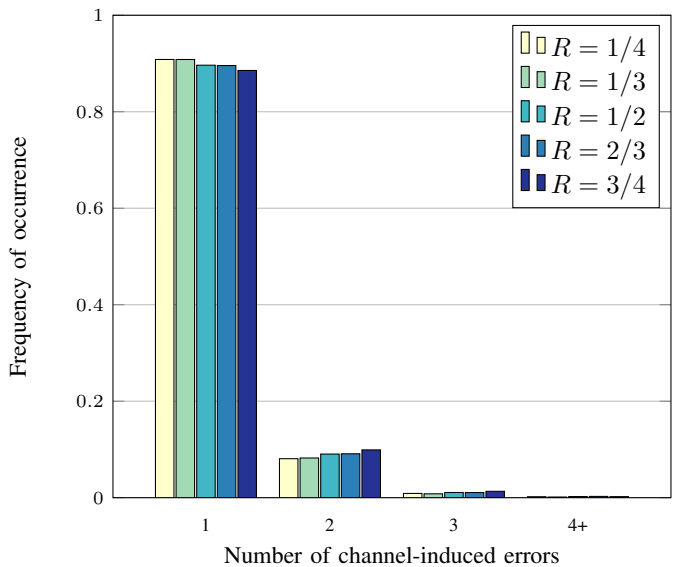
\begin{figure}
   \centering
   \begin{tikzpicture}

 \definecolor{r1}{RGB}{255,255,204}
 \definecolor{r2}{RGB}{161,218,180}
 \definecolor{r3}{RGB}{65,182,196}
 \definecolor{r4}{RGB}{44,127,184}
 \definecolor{r5}{RGB}{37,52,148}

  \pgfplotsset{
    label style = {font=\fontsize{9pt}{7.2}\selectfont},
    tick label style = {font=\fontsize{7pt}{7.2}\selectfont}
  }

\begin{axis}[
	width  = 9cm,
        height = 8cm,
        major x tick style = transparent,
        ybar=2*\pgflinewidth,
        bar width=7pt,
        ymajorgrids = true,
        xlabel = {Number of channel-induced errors},
        ylabel = {Frequency of occurrence},
        symbolic x coords={1,2,3,4+},
        xtick = data,
        scaled y ticks = false,
        enlarge x limits=0.25,
        ymin=0,
        ymax=1,
]

	\addplot[style={black,fill=r1,mark=none}]
            coordinates {(1,0.908356) (2,0.0808081) (3,0.00881543) (4+,0.0020202)};
            \addlegendentry{$R = 1/4 $}

	\addplot[style={black,fill=r2,mark=none}]
            coordinates {(1,0.908329) (2,0.0822644) (3,0.00803831) (4+,0.00136822)};
            \addlegendentry{$R = 1/3 $}

        \addplot[style={black,fill=r3,mark=none}]
              coordinates {(1,0.896477) (2,0.090411) (3,0.0107632) (4+,0.00234834)};
            \addlegendentry{$R = 1/2 $}
            
        \addplot[style={black,fill=r4,mark=none}]
              coordinates {(1,0.895787) (2,0.0909831) (3,0.0105842) (4+,0.00264604)};
            \addlegendentry{$R = 2/3 $}

        \addplot[style={black,fill=r5,mark=none}]
              coordinates {(1,0.885592) (2,0.0990347) (3,0.0132285) (4+,0.00214516)};
            \addlegendentry{$R = 3/4 $}

\end{axis}
\end{tikzpicture}
   \caption{Frequency of occurrence of channel-incurred errors for polar code with $N = 512$, $R \in \{\frac{1}{4},\frac{1}{3},\frac{1}{2},\frac{2}{3},\frac{3}{4}\}$ at target FER $=10^{-4}$.}
   \label{fig:errorlikelihoods-rate}
 \end{figure}

The SC-Flip algorithm was first introduced in \cite{SCFlip14} to improve finite-length error-correction performance of polar codes. It was observed that when SC decoding fails, it is due to one or more incorrect bit estimations at one of the leaf nodes, that are in fact propagated through the tree due to sequential nature of SC, resulting in more incorrect estimations. As a result, incorrect bit decisions are classified into two categories: errors due to channel noise, and errors caused by a previous incorrect estimation. Consequently, the first observed error is always due to channel noise. Fig.~\ref{fig:errorlikelihoods-snr} depicts the frequency of occurrence for channel-induced errors, for $PC(512,256)$ with different $E_b/N_0$ points. It can be observed that most of the decoding failures are due to a single channel-induced error, whose  frequency increases with $E_b/N_0$. On the other hand, Fig.~\ref{fig:errorlikelihoods-rate} highlights that failures due to a single error are more prominent at target FER $=10^{-4}$, and is observed to be around $90\%$ for code rates $R \in \{\frac{1}{4},\frac{1}{3},\frac{1}{2},\frac{2}{3},\frac{3}{4}\}$. Throughout the rest of this work, the occurrence of a single channel-induced error is denoted as $E_1$.

In order to observe the impact of $E_1$ on the error-correction performance, a genie-like decoder called SC-Oracle was introduced in \cite{SCFlip14}: it has foreknowledge of the transmitted codeword, which it uses to identify the first channel-induced error and ensure that SC estimates the bit correctly. In this work, we target the correction of $E_1$, and thus consider SC-Oracle correcting only $E_1$ as a baseline. We refer to the corresponding frame error rate (FER) performance as SCO-1 performance. Original SC-Flip algorithm \cite{SCFlip14} uses a cyclic-redundancy check (CRC) code with a $C$-bit remainder to encode the information bits. If the CRC is successful at the end of SC decoding, the estimated codeword is assumed to be correct. In case the CRC fails, the $T_{max}-1$ LLRs with the smallest magnitude, representing the bit estimations with lowest reliability, are stored and sorted, after which a set of SC decoding attempts are initiated. At each decoding attempt, one of the selected bits is flipped, in ascending reliability order, and the CRC is checked at the end of each iteration. This process continues for $T_{max}$ attempts, or until the CRC passes. It should be noted that the initial SC decoding is counted as the first iteration towards $T_{max}$ maximum iterations.

Since decoding of a single codeword can take up to $T_{max}$ iterations, the decoding latency of SC-Flip is not fixed. As $T_{max}$ increases, the error-correction performance of SC-Flip decoding improves towards its lower bound, which is the SCO-1 performance with $K+C$ non-frozen bits. The performance gap between SC-Flip and SCO-1 is due to two possible cases: either the estimated codeword with a successful CRC check results in undetected errors, or the decoding stopped after reaching $T_{max}$ iterations without correcting the error.

Improvements for SC-Flip have been proposed recently in \cite{SCFlip17-conf} and \cite{SCFlip17-jour}, where a generalized SC-Flip decoding algorithm is used to correct more than one erroneous hard decision through nested flips. A simulation-based scaling metric is also introduced in order to help the SC-Flip decoder detect the erroneous bit indices more accurately. Simulation results show an improvement of $0.4$~dB in error-correction performance when $T_{max}=10$ with respect to SC-Flip. Practical implementation of this approach requires either a $T_{max}$ value that is larger by an order of magnitude, or parallel SC-Flip decoders similar to SC-List decoding. Partitioned SC-Flip (PSCF) decoding was introduced in \cite{PSCF-ICC18}, in which the CRC is distributed into multiple partitions in the codeword, in order to improve the error-correction performance and reduce average number of iterations significantly. The first hardware implementation of baseline SC-Flip decoder was recently implemented in a multi-decoder chip with 28 nm CMOS technology which supports configurable CRC length and $T_{max}$ \cite{PolarBear}. Results show that SC-Flip decoding with $C=8$ and $T_{max} = 8$ is up to $32.4\%$ faster than SC-List at target FER of $10^{-2}$, while its energy consumption is similar to that of an SC decoder.

\subsection{Improving SC-Flip Based on Error Distribution}\label{sec:WCNC}

In \cite{SCF-WCNC18} we performed a simulation campaign with SC-Oracle, to identify the bit indices more likely to incur in $E_1$. We observed that, although all non-frozen indices are theoretically susceptible to channel-induced errors, a fraction of them is more likely to incur one than the others. Thus, only those that bring a substantial contribution to the FER are considered as possibly erroneous. These observations allow to reduce the set of bits to be flipped, and two bit-flipping index selection criteria for SC-Flip have been proposed. The first one, called fixed-index selection (FIS), replaces the selection and sorting process of SC-Flip with a list of critical indices that is pre-sorted based on their $E_1$ occurrence frequency. The second method, called enhanced index selection (EIS), applies the LLR-based index selection of SC-Flip to the critical set only, thus excluding indices with a low probability of $E_1$. Simulations have shown that there is no error-correction performance loss when FIS is applied to SC-Flip for low-rate polar codes, while the LLR sorting needed by the original index selection is avoided, which represents an estimated $24.6\%$ of the total logic complexity in ASIC implementations. Simulation results for the EIS criterion show gains with respect to SC-Flip of up to $0.4$ dB and $0.42$ dB for $PC(1024,170)$ and $PC(1024,256)$ respectively, that decrease at higher code rates.

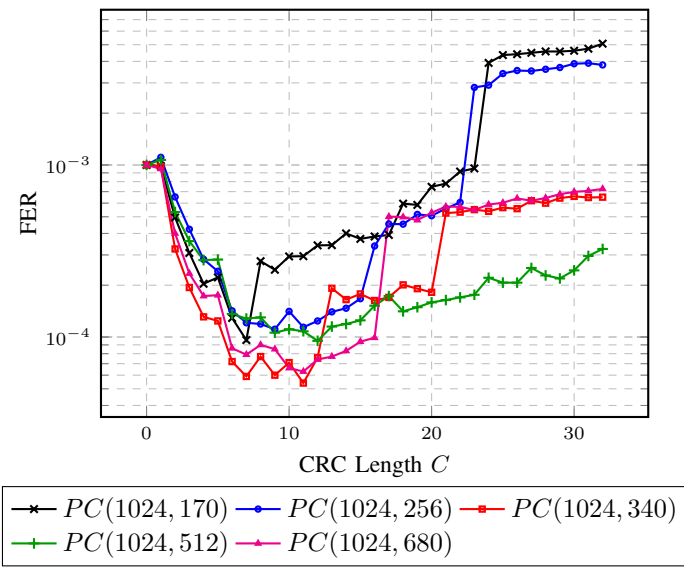
\begin{figure}[t]
  \centering
   \scalebox{1}{\begin{tikzpicture}[spy using outlines=
	{rectangle, magnification=2, connect spies}]
  \pgfplotsset{
    label style = {font=\fontsize{9pt}{7.2}\selectfont},
    tick label style = {font=\fontsize{7pt}{7.2}\selectfont}
  }

\begin{axis}[
	scale = 1,
    ymode=log,
    xlabel={CRC Length $C$}, xlabel style={yshift=0.4em},
    ylabel={FER}, ylabel style={yshift=-0.75em},
    grid=both,
    ymajorgrids=true,
    xmajorgrids=true,
    grid style=dashed,
    width=\columnwidth, height=7cm,
    thick,
    mark size=3,
    legend to name=CRC_vs_FER3,
    legend columns=3,
]

\addplot[
    color=black,
    mark=x,
    thick,
    mark size=2,
]
table {
0	1.00000e-03
1	9.85000e-04
2	4.95000e-04
3	3.08000e-04
4	2.04000e-04 
5	2.22000e-04
6	1.29000e-04
7	9.60000e-05
8	2.76000e-04
9	2.46000e-04
10	2.94000e-04
11	2.95000e-04
12	3.41000e-04
13	3.42000e-04
14	4.00000e-04
15	3.72000e-04
16	3.84000e-04
17	3.92000e-04
18	5.96000e-04
19	5.86000e-04
20	7.47000e-04
21	7.79000e-04
22	9.16000e-04
23	9.55000e-04
24	3.92100e-03
25	4.35500e-03
26	4.40600e-03
27	4.48800e-03
28	4.57100e-03
29	4.56600e-03
30	4.61400e-03
31	4.74200e-03
32	5.07200e-03
};
\addlegendentry{$PC(1024,170)$}	

\addplot[
    color=blue,
    mark=o,
    thick,
    mark size=1,
]
table {
0	1.00000e-3
1	1.10800e-3
2	6.52000e-4
3	4.23000e-4
4	2.83000e-4
5	2.41000e-4
6	1.42000e-4
7	1.21000e-4
8	1.19000e-4
9	1.11000e-4
10	1.41000e-4
11	1.14000e-4
12	1.24000e-4
13	1.40000e-4
14	1.47000e-4
15	1.67000e-4
16	3.38000e-4
17	4.54000e-4
18	4.53000e-4
19	5.16000e-4
20	5.07000e-4
21	5.58000e-4
22	6.06000e-4
23	2.82300e-3
24	2.91100e-3
25	3.40000e-3
26	3.54000e-3
27	3.52000e-3
28	3.60100e-3
29	3.68600e-3
30	3.87800e-3
31	3.90700e-3
32	3.81600e-3
};
\addlegendentry{$PC(1024,256)$}	

\addplot[
    color=red,
    mark=square,
    thick,
    mark size=1,
]
table {
0	1.00000e-3
1	9.79000e-4
2	3.25000e-4
3	1.94000e-4
4	1.31000e-4
5	1.24000e-4
6	7.20000e-5
7	5.90000e-5
8	7.70000e-5
9	6.00000e-5
10	7.10000e-5
11	5.40000e-5
12	7.60000e-5
13	1.92000e-4
14	1.65000e-4
15	1.78000e-4
16	1.63000e-4
17	1.70000e-4
18	2.01000e-4
19	1.91000e-4
20	1.82000e-4
21	5.25000e-4
22	5.32000e-4
23	5.50000e-4
24	5.38000e-4
25	5.65000e-4
26	5.57000e-4
27	6.20000e-4
28	6.00000e-4
29	6.42000e-4
30	6.57000e-4
31	6.47000e-4
32	6.50000e-4
};
\addlegendentry{$PC(1024,340)$}	

\addplot[
    color=green!60!black,
    mark=+,
    thick,
    mark size=2,
]
table {
0	1.00000e-03
1	1.07000e-03
2	5.33000e-04
3	3.61000e-04
4	2.80000e-04
5	2.82000e-04
6	1.36000e-04
7	1.28000e-04
8	1.30000e-04
9	1.06000e-04
10	1.11000e-04
11	1.08000e-04
12	9.50000e-05
13	1.15000e-04
14	1.19000e-04
15	1.25000e-04
16	1.52000e-04
17	1.73000e-04
18	1.41000e-04
19	1.49000e-04
20	1.59000e-04
21	1.64000e-04
22	1.70000e-04
23	1.76000e-04
24	2.21000e-04
25	2.07000e-04
26	2.07000e-04
27	2.52000e-04
28	2.28000e-04
29	2.19000e-04
30	2.44000e-04
31	2.96000e-04
32	3.25000e-04
};
\addlegendentry{$PC(1024,512)$}	

\addplot[
    color=magenta,
    mark=triangle,
    thick,
    mark size=1,
]
table {
0	1.00000e-3
1	9.49000e-4
2	4.00000e-4
3	2.34000e-4
4	1.73000e-4
5	1.75000e-4
6	8.60000e-5
7	7.90000e-5
8	9.00000e-5
9	8.50000e-5
10	6.60000e-5
11	6.30000e-5
12	7.40000e-5
13	7.70000e-5
14	8.30000e-5
15	9.40000e-5
16	9.90000e-5
17	5.01000e-4
18	4.98000e-4
19	4.78000e-4
20	5.28000e-4
21	5.73000e-4
22	5.71000e-4
23	5.47000e-4
24	5.89000e-4
25	6.03000e-4
26	6.39000e-4
27	6.20000e-4
28	6.43000e-4
29	6.76000e-4
30	6.98000e-4
31	7.07000e-4
32	7.26000e-4
};
\addlegendentry{$PC(1024,680)$}	

\end{axis}
\end{tikzpicture}}
   \ref{CRC_vs_FER3}
   \\
  \caption{Error correction performance of SC-Flip decoding with respect to CRC length $C$, for polar codes of various rates.}
  \label{fig:CRC_Sweep_ALL}
\end{figure}

A similar approach has been recently taken in \cite{SCF-GLOBECOM17} to improve the error-correction performance of SC-Flip. A set of indices is created from each sub-tree of the polar code that is a Rate-1 node (i.e. nodes without any frozen bits), including almost all channel-induced errors. The size of the index set increases with $E_b/N_0$, and can be used progressively to correct multiple errors, achieving an error-correction performance comparable to that of SC-List with $L=32$. However, a very high $T_{max}$ is needed by this technique, since the bit flipping search space must include the entire set; which also leads to very long worst-case latency. 

\section{Thresholded Index Selection for SC-Flip Decoding} \label{sec:ourtech}

The FIS and EIS criteria \cite{SCF-WCNC18} observe the distribution of $E_1$ occurrences to limit the set of indices for bit flipping to reduce the implementation complexity and improve the error-correction performance of SC-Flip. However, their benefits diminish at higher code rates. In this Section, we use the SC-Oracle decoder to provide further insight about failed decoding attempts, and propose a new index selection scheme for SC-Flip. We detail the proposed method focusing on low-rate codes, and extend it to higher-rate polar codes in Section \ref{sec:highrate}. 
The CRC remainder length for each rate is selected with respect to the FER observation in Fig.~\ref{fig:CRC_Sweep_ALL}; it can be seen that the best $C$ value is different for each rate. 
The polar code used in this work is constructed targeting an $E_b/N_0$ value of $2.5$ dB, using \cite{pc-construction}. The distribution of $E_1$, the critical set, and the optimal threshold values may change with a different polar code construction, but the proposed approach is independent of it.

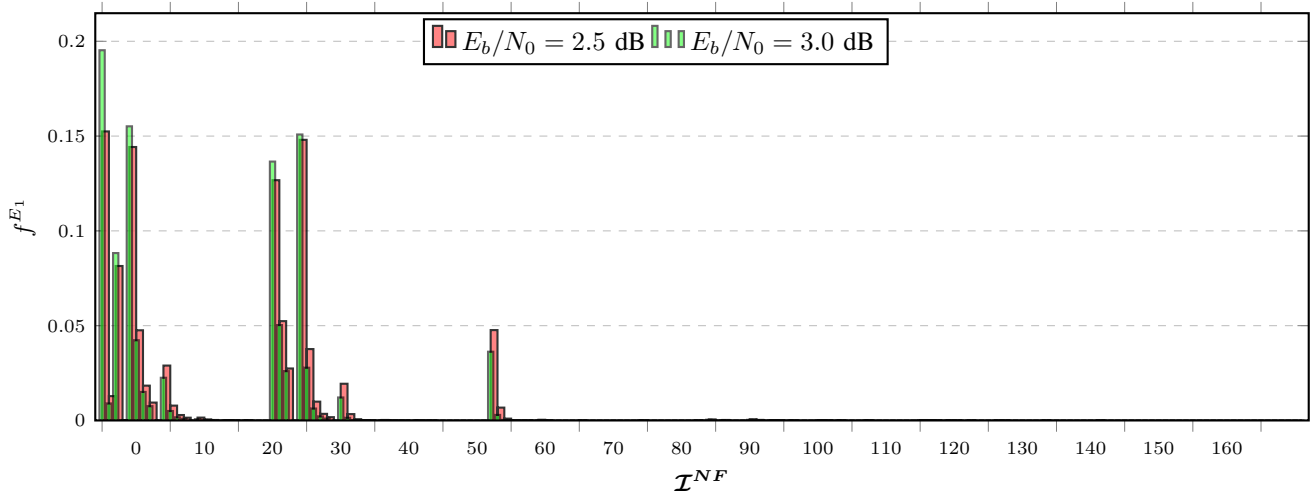
\begin{figure*}[t]
  \centering
   \scalebox{1}{\begin{tikzpicture}
  \pgfplotsset{
    label style = {font=\fontsize{9pt}{7.2}\selectfont},
    tick label style = {font=\fontsize{7pt}{7.2}\selectfont},
    yticklabel style={/pgf/number format/fixed}
  }

\begin{axis}[
	scale = 1,
    ybar interval,
    xlabel={$\boldsymbol{\mathcal{I}^{NF}}$}, xlabel style={yshift=0em},
    ylabel={$f^{E_1}$}, ylabel style={yshift=-0.75em},
    grid=both,
    xmin = -1,
    xmax = 177,
    ymin = 0,
    xtick={0,10,20,30,40,50,60,70,80,90,100,110,120,130,140,150,160,170,180},
    ymajorgrids=true,
    xmajorgrids=false,
    grid style=dashed,
    width=2*\columnwidth, height=7cm,
    thick,
    mark size=3,
    legend style={at={(0.27,0.93)},anchor=west},
    legend columns=2
]

\addplot[
    bar width=5pt,
    fill=red!60,opacity=0.8,
    thick,
]
table {
0	0.1524547021
1	0.0128292027
2	0.0814661946
3	8.89865005585323E-05
4	0.1442300774
5	0.0475604446
6	0.0183615124
7	0.0094268891
8	0
9	0.0289471193
10	0.0078213454
11	0.0028627147
12	0.0014578639
13	0
14	0.0015051972
15	0.000545279
16	0.0002404529
17	0
18	0.0001703997
19	8.51998409602969E-05
20	0
21	3.78665959823542E-05
22	0
23	0
24	0
25	0.1267300301
26	0.0523846489
27	0.0274097355
28	0
29	0.1480186304
30	0.0376488631
31	0.0099191548
32	0.0034818335
33	0.0017286101
34	0
35	0.0193574039
36	0.0033455138
37	0.0006626654
38	0.000183653
39	0.0001003465
40	0
41	0.0002555995
42	3.59732661832365E-05
43	0.00001704
44	9.46664899558854E-06
45	0
46	1.89332979911771E-06
47	3.78665959823542E-06
48	3.78665959823542E-06
49	0
50	0
51	0
52	0
53	0
54	0
55	0
56	0
57	0.047666471
58	0.0067705474
59	0.0009618115
60	0.0001760797
61	5.30132343752958E-05
62	2.65066171876479E-05
63	0
64	0.0003635193
65	2.08266277902948E-05
66	0
67	0
68	0
69	0
70	1.89332979911771E-06
71	0
72	0
73	0
74	0
75	0
76	0
77	0
78	0
79	0.0001003465
80	0
81	0
82	0
83	0
84	0
85	0
86	0
87	1.89332979911771E-06
88	0
89	0.0005490656
90	0
91	0.0002537062
92	0
93	0
94	0
95	0.0006437321
96	0.0002877861
97	0
98	0.000160933
99	8.14131813620615E-05
100	0
101	4.35465853797073E-05
102	0
103	0
104	0
105	3.78665959823542E-05
106	2.46132873885302E-05
107	0
108	0.00001136
109	0
110	0
111	0
112	1.89332979911771E-06
113	0
114	0
115	0
116	0
117	0
118	0
119	0
120	0.0002158396
121	6.24798833708844E-05
122	4.16532555805896E-05
123	0
124	2.08266277902948E-05
125	9.46664899558854E-06
126	0
127	0.00000568
128	0
129	0
130	0
131	9.46664899558854E-06
132	1.89332979911771E-06
133	0
134	0
135	0
136	0
137	0
138	1.89332979911771E-06
139	0
140	0
141	0
142	0
143	0
144	0
145	0
146	1.89332979911771E-06
147	0
148	0
149	0
150	0
151	0
152	0
153	0
154	0
155	0
156	0
157	0
158	0
159	0
160	0
161	0
162	0
163	0
164	0
165	0
166	0
167	0
168	0
169	0
170	0
171	0
172	0
173	0
174	0
175	0
176	0

};
\addlegendentry{$E_b/N_0 = 2.5$ dB}

\addplot[
    ybar,
    bar width=2pt,
    fill=green!80,opacity=0.6,
    thick,
    mark size=3,
]
table {
0	0.1952850251
1	0.0089781463
2	0.0882920008
3	2.48243675992354E-05
4	0.1551440227
5	0.042317272
6	0.0150435668
7	0.007538333
8	0
9	0.0225157014
10	0.0049896979
11	0.0016549578
12	0.0008440285
13	0
14	0.0007033571
15	0.0002896176
16	0.0001241218
17	0
18	6.61983135979611E-05
19	2.48243675992354E-05
20	0
21	4.13739459987257E-05
22	0
23	0
24	0
25	0.1365505714
26	0.0503189931
27	0.026024212
28	0
29	0.1508411323
30	0.0278115665
31	0.006280565
32	0.0021431704
33	0.0010095243
34	0
35	0.0121225662
36	0.0013901646
37	0.0002234193
38	7.44731027977062E-05
39	4.13739459987257E-05
40	0
41	9.92974703969416E-05
42	8.27478919974514E-06
43	0
44	0
45	0
46	8.27478919974514E-06
47	0
48	0
49	0
50	0
51	0
52	0
53	0
54	0
55	0
56	0
57	0.0362601263
58	0.0029210006
59	0.0002978924
60	2.48243675992354E-05
61	0
62	0
63	0
64	9.10226811971965E-05
65	1.65495783994903E-05
66	0
67	0
68	0
69	0
70	0
71	0
72	0
73	0
74	0
75	0
76	0
77	0
78	0
79	6.61983135979611E-05
80	0
81	0
82	0
83	0
84	0
85	0
86	0
87	0
88	0
89	0.0003971899
90	0
91	0.0001654958
92	0
93	0
94	0
95	0.0003309916
96	0.0001323966
97	0
98	5.79235243982159E-05
99	4.96487351984708E-05
100	0
101	1.65495783994903E-05
102	0
103	0
104	0
105	6.61983135979611E-05
106	2.48243675992354E-05
107	0
108	8.27478919974514E-06
109	0
110	0
111	0
112	8.27478919974514E-06
113	0
114	0
115	0
116	0
117	0
118	0
119	0
120	0.0001406714
121	5.79235243982159E-05
122	3.30991567989805E-05
123	0
124	8.27478919974514E-06
125	0
126	0
127	0
128	0
129	0
130	0
131	0
132	0
133	0
134	0
135	0
136	0
137	0
138	0
139	0
140	0
141	0
142	0
143	0
144	0
145	0
146	0
147	0
148	0
149	0
150	0
151	0
152	0
153	0
154	0
155	0
156	0
157	0
158	0
159	0
160	0
161	0
162	0
163	0
164	0
165	0
166	0
167	0
168	0
169	0
170	0
171	0
172	0
173	0
174	0
175	0
176	0

};
\addlegendentry{$E_b/N_0 = 3.0$ dB}
\end{axis}
\end{tikzpicture}}
   \\
  \caption{$f^{E_1}$ values corresponding to each non-frozen index of $PC(1024,170)$, $C = 7$, ${\rm FER_t}=10^{-4}$.}
  \label{fig:spikes}
\end{figure*}

As mentioned in Section~\ref{sec:WCNC}, all non-frozen indices are theoretically susceptible to channel-induced errors. If the maximum number of iterations is large enough to flip all possibly erroneous bits, i.e. $T_{max} = K+C$, and assuming a perfect CRC, SC-Flip can identify and correct all $E_1$ occurrences and achieve the SCO-1 performance. Let us define the set of non-frozen indices as $\boldsymbol{\mathcal{I}^{NF}}$, and their associated frequency of $E_1$ occurrence as $\boldsymbol{f^{E_1}}$. For example, consider the polar code $PC(8,5)$ in Fig.~\ref{fig:scdecode}. The non-frozen indices are $\boldsymbol{\mathcal{I}^{NF}} = \{u_3, u_4, u_5, u_6, u_7\}$, and assume $\boldsymbol{f^{E_1}} = \{ 0.25, 0.40, 0.20, 0.13, 0.02\}$, respectively. In this situation, the most critical index is $u_4$, followed by $u_3$, and so on. Note that the summation of all the elements in $\boldsymbol{f^{E_1}}$ is equal to 1. Considering the whole $\boldsymbol{\mathcal{I}^{NF}}$ for bit-flipping is not feasible for practical applications since the maximum and average number of iterations would be extremely large. Fig.~\ref{fig:spikes} depicts $\boldsymbol{f^{E_1}}$ for all non-frozen bits in $PC(1024,170)$ with $C=32$ at two different $E_b/N_0$ values. As the simulations to identify the frequency of $E_1$ occurrence have to run for a finite time, a target FER ${\rm (FER_t})$ is required to determine the minimum amount of simulated frames. In this work, we set ${\rm FER_t}$ as $10^{-4}$. It can be seen that with the target ${\rm FER_t}$, only a limited set of non-frozen indices has non-negligible $\boldsymbol{f^{E_1}}$. Let us define the FER obtained with SCO-1 at a certain $E_b/N_0$ as ${\rm FER_{SCO-1}}$ and the FER obtained with SC at the same $E_b/N_0$ point (${\rm FER_{SC}}$). Then:
\begin{equation*}
  \sum f^{E_1}_i\times ({\rm FER_{SC}}-{\rm FER_{SCO-1}})={\rm FER_{SC}}-{\rm FER_{SCO-1}}~,
\end{equation*}
where $f^{E_1}_i$ is the $i^{\rm th}$ element of $\boldsymbol{f^{E_1}}$. This shows that considering separately all the contributions to $\text{FER}_{\text{SC}}$ of $E_1$ and summing them together, we get as a result the difference between $\text{FER}_{\text{SC}}$ and $\text{FER}_{\text{SCO-1}}$. The critical set $\boldsymbol{\mathcal{I}^{C}}$ is defined as the smallest subset of $\boldsymbol{\mathcal{I}^{NF}}$ that satisfies the following:
\begin{equation*}
  \sum f^{E_1}_i\times ({\rm FER_{SC}}-{\rm FER_{SCO-1}})\ge\gamma\times({\rm FER_{SC}}-{\rm FER_{SCO-1}})~,
\end{equation*}
where $\gamma \approx 1$. This means that if we exclude some of the non-frozen bit indices from the previous calculation, we can identify a factor $\gamma\approx 1$ so that the summation of the considered $f_i^{E_1}$ is equal to $\gamma$, and the consequent difference in $\text{FER}_{\text{SC}}-\text{FER}_{\text{SCO-1}}$, and thus FER degradation, is negligible. In the remainder of this work, $\gamma$ is set to $0.9999$. The non-frozen indices belonging to $\boldsymbol{\mathcal{I}^{C}}$ are called critical indices. 

\begin{figure}
  \centering
   \scalebox{1}{\begin{tikzpicture}
  \pgfplotsset{
    label style = {font=\fontsize{9pt}{7.2}\selectfont},
    tick label style = {font=\fontsize{7pt}{7.2}\selectfont},
    yticklabel style={/pgf/number format/fixed}
  }

\begin{axis}[
	scale = 1,
    xlabel={$\boldsymbol{\mathcal{I}^{C}}$}, xlabel style={yshift=0em},
    ylabel={Normalized average LLR magnitude}, ylabel style={yshift=-0.75em},
    grid=both,
    xmin = -1,
    xmax = 70,
    ymin = 0,
    ymajorgrids=true,
    xmajorgrids=true,
    grid style=dashed,
    width=\columnwidth, height=7cm,
    thick,
    mark size=3,
    legend style={
      anchor={center},
      cells={anchor=west},
      column sep= 2mm,
      font=\fontsize{7pt}{7.2}\selectfont,
    },
    legend to name=LLRdist-suspected,
    legend columns=2,
]

\addplot[
    color=blue,
    thick,
    mark size=4,
]
table {
0	0.4702348961
1	0.3173943283
2	0.3206857548
3	0.329309503
4	0.3595540943
5	0.3818985594
6	0.3998854851
7	0.3895194236
8	0.4024866604
9	0.428440454
10	0.4206597606
11	0.4489811331
12	0.4449818919
13	0.6607670394
14	0.4796894046
15	0.4853431409
16	0.5052620671
17	0.5174635291
18	0.5384836957
19	0.5520498526
20	0.5638895036
21	0.5805855016
22	0.6043455963
23	0.6060141422
24	0.6299270406
25	0.638011585
26	0.5134800951
27	0.5114005599
28	0.6637194876
29	0.6921620492
30	0.5436175482
31	0.6922762129
32	0.5470617784
33	0.7062076936
34	0.5671774174
35	0.7005451754
36	0.7205396253
37	0.7357321755
38	0.5641090491
39	0.5992907801
40	0.7450215857
41	0.8983697428
42	0.7782713161
43	0.5938144366
44	0.6101960454
45	0.7857780166
46	0.6294510659
47	0.6408604077
48	0.8294851394
49	0.6233582386
50	0.7815065389
51	0.8321267111
52	0.6530934842
53	0.8168006773
54	0.6614976869
55	0.8454241444
56	0.6886581026
57	0.6916579727
58	0.7217620548
59	0.8907980568
60	0.7277758458
61	1
62	0.954497873
63	0.7304086357
64	0.9747487521
65	0.7905114181
66	0.8905117694
67	0.751774806
68	0.8297222486
69	0.9324853607

};
\addlegendentry{Correct bit}

\addplot[
    color=red,
    thick,
    mark size=4,
]
table {
0	0.0649698432
1	0.042937484
2	0.0435580074
3	0.0441885422
4	0.0462020381
5	0.0447280973
6	0.0421657376
7	0.0433335441
8	0.0409015066
9	0.0423095838
10	0.046644115
11	0.040868487
12	0.0431570295
13	0.0561829288
14	0.0406858251
15	0.0440942254
16	0.0418290425
17	0.039147777
18	0.0422474085
19	0.0400113812
20	0.0422453008
21	0.0397925383
22	0.0440603276
23	0.0457283467
24	0.0394493447
25	0.0368557217
26	0.0343601038
27	0.034432466
28	0.038547803
29	0.037208224
30	0.0305089241
31	0.0378343678
32	0.0359722705
33	0.0435054922
34	0.0355452984
35	0.0423248641
36	0.029828333
37	0.0449364021
38	0.0359569902
39	0.0360247858
40	0.053201325
41	0.0954092153
42	0.049684557
43	0.0328568317
44	0.0287283221
45	0.0376764707
46	0.0363096681
47	0.0300210061
48	0.0464247451
49	0.0314956495
50	0.0397282553
51	0.0396495702
52	0.0370875618
53	0.0384703473
54	0.034566301
55	0.0194088781
56	0.0434684329
57	0.0537101437
58	0.0241201494
59	0.0531064813
60	0.0204533879
61	0.0082804141
62	0.0640821768
63	0.1284587202
64	0.0249459918
65	0.0220163763
66	0.025937635
67	0.0337388778
68	0.0949351726
69	0.0183174734
};
\addlegendentry{$E_1$}

\end{axis}
\end{tikzpicture}}
   \ref{LLRdist-suspected}
  \vspace{2pt}
  \caption{Normalized average LLR magnitudes for $\boldsymbol{\mathcal{I}^{C}}$, $PC(1024,170)$, $C=7$ at $E_b/N_0 = 2.5$ dB. Correct bit estimation is in blue, $E_1$ occurrence is in red.}
  \label{fig:avgLLRs-TIS}
\end{figure}
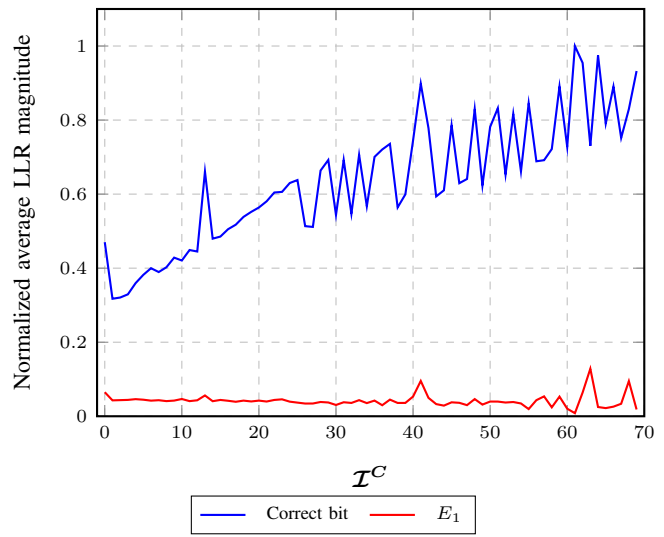

Additional insight on the behavior of erroneous indices can be gained by observing the magnitude of LLRs. Fig.~\ref{fig:avgLLRs} plots the average LLR magnitude for $\boldsymbol{\mathcal{I}^{NF}}$ of $PC(1024,170)$, normalized with the maximum LLR value: it can be seen that all of the error-prone indices identified in Fig.~\ref{fig:spikes}, highlighted in red in here, are associated to small average LLR magnitudes. Fig.~\ref{fig:avgLLRs-TIS} depicts an example of average LLR magnitudes for the indices in the critical set $\boldsymbol{\mathcal{I}}^C$, for the case in which an $E_1$ occurs at the corresponding index (in red) and for when it does not (in blue). The average LLR magnitudes are substantially larger in case of successful decoding, compared to when it fails. 

\begin{figure}
  \centering
   \scalebox{1}{\begin{tikzpicture}
  \pgfplotsset{
    label style = {font=\fontsize{9pt}{7.2}\selectfont},
    tick label style = {font=\fontsize{7pt}{7.2}\selectfont},
    yticklabel style={/pgf/number format/fixed}
  }

\begin{axis}[
    scale = 1,
    ybar interval,
    xlabel={Non-frozen indices}, xlabel style={yshift=0em},
    ylabel={Normalized average LLR magnitude}, ylabel style={yshift=-0.75em},
    grid=both,
    xmin = -1,
    xmax = 180,
    ymin = 0,
    xtick={0,20,40,60,80,100,120,140,160,180},
    ymajorgrids=true,
    xmajorgrids=false,
    grid style=dashed,
    width=\columnwidth, height=7cm,
    thick,
    mark size=3,
    legend style={
      anchor={center},
      cells={anchor=west},
      column sep= 2mm,
      font=\fontsize{9pt}{7.2}\selectfont,
    },
    legend to name=LLRdist,
    legend columns=2,
]

\addplot[
    color=white!25!black,
    thick,
    mark size=5,
]
table {
0	0.0220899167
1	0.0310404205
2	0.0168905372
3	0.042202127
4	0.0150646447
5	0.0182981989
6	0.0209036229
7	0.0227996469
8	0.0546876676
9	0.0201265666
10	0.0237353652
11	0.0264894679
12	0.0284683295
13	0.0664030825
14	0.0283899473
15	0.0311791157
16	0.0331750563
17	0.0759143076
18	0.0345620086
19	0.0365603419
20	0.0826973375
21	0.0389661801
22	0.0875174297
23	0.0909720217
24	0.1983886272
25	0.0154697569
26	0.0179401985
27	0.0197610581
28	0.048337555
29	0.0149100255
30	0.0189073522
31	0.022534055
32	0.0252959959
33	0.0272737849
34	0.0640755439
35	0.0210914926
36	0.0259332844
37	0.0299714524
38	0.0329090519
39	0.0349983911
40	0.0800700489
41	0.0325206063
42	0.0367123209
43	0.0397149363
44	0.0418464369
45	0.0939744722
46	0.0418329882
47	0.0448388215
48	0.0469763451
49	0.1042681166
50	0.048523032
51	0.0506594005
52	0.1116360426
53	0.0532408148
54	0.1168010165
55	0.1204775538
56	0.2586661826
57	0.0187851585
58	0.0243085453
59	0.02959167
60	0.0338483181
61	0.0369129793
62	0.0390902715
63	0.0886552091
64	0.0325152433
65	0.0383703105
66	0.042839168
67	0.0460132342
68	0.0482612355
69	0.1073753517
70	0.0457901337
71	0.0503004101
72	0.0534895752
73	0.0557525928
74	0.1224247325
75	0.0557749524
76	0.0589519063
77	0.0612112111
78	0.1333223323
79	0.0281524905
80	0.0628345476
81	0.0650889844
82	0.1410541167
83	0.0678080214
84	0.1464814647
85	0.1503403437
86	0.3192630424
87	0.0438047541
88	0.0501226062
89	0.0240237292
90	0.0547724029
91	0.0256989629
92	0.0580714681
93	0.0604159207
94	0.1320970949
95	0.0241214181
96	0.0255371655
97	0.0579160238
98	0.0264997814
99	0.0278952319
100	0.0625438734
101	0.0295693105
102	0.0658266021
103	0.0681616488
104	0.1475491126
105	0.0292830917
106	0.0306799449
107	0.068130461
108	0.0323506407
109	0.07138951
110	0.073710943
111	0.1585853252
112	0.0343119281
113	0.0753337844
114	0.0776443264
115	0.1664020924
116	0.0804260691
117	0.1719292745
118	0.1758714862
119	0.3705992525
120	0.0266439221
121	0.02866478
122	0.0301052797
123	0.0671986205
124	0.0310747436
125	0.0324915636
126	0.0718609582
127	0.0341882493
128	0.0751738022
129	0.0775283207
130	0.1664161187
131	0.0339057434
132	0.0353156327
133	0.0775023308
134	0.0370034076
135	0.0807808516
136	0.0831107004
137	0.1774853343
138	0.0389773187
139	0.0847352745
140	0.0870504369
141	0.1852930256
142	0.0898400178
143	0.1908202078
144	0.1947492182
145	0.4084075214
146	0.0371353372
147	0.0385666785
148	0.0841139925
149	0.0402745027
150	0.0874159454
151	0.0897641109
152	0.1908853887
153	0.0422719284
154	0.0913961106
155	0.093721999
156	0.1986988556
157	0.096523131
158	0.2042235625
159	0.2081533981
160	0.4352109306
161	0.0445624211
162	0.0960858409
163	0.0984274057
164	0.2082004274
165	0.1012384386
166	0.2137201838
167	0.2176442439
168	0.4541926222
169	0.1045544179
170	0.2204346499
171	0.224359535
172	0.4676124784
173	0.2290962946
174	0.4770744466
175	0.4838219157
176	1
177	0
};

\addplot[
    color=red,
    thick,
    mark size=3,
]
table {
0	0.0220899167
1	0.0310404205
2	0.0168905372
3	0.042202127
4	0.0150646447
5	0.0182981989
6	0.0209036229
7	0.0227996469
8	0
9	0.0201265666
10	0.0237353652
11	0.0264894679
12	0.0284683295
13	0
14	0.0283899473
15	0.0311791157
16	0.0331750563
17	0
18	0.0345620086
19	0.0365603419
20	0
21	0.0389661801
22	0
23	0
24	0
25	0.0154697569
26	0.0179401985
27	0.0197610581
28	0
29	0.0149100255
30	0.0189073522
31	0.022534055
32	0.0252959959
33	0.0272737849
34	0
35	0.0210914926
36	0.0259332844
37	0.0299714524
38	0.0329090519
39	0.0349983911
40	0
41	0
42	0.0367123209
43	0
44	0
45	0
46	0
47	0
48	0
49	0
50	0
51	0
52	0
53	0
54	0
55	0
56	0
57	0.0187851585
58	0.0243085453
59	0.02959167
60	0.0338483181
61	0.0369129793
62	0.0390902715
63	0
64	0.0325152433
65	0.0383703105
66	0
67	0
68	0
69	0
70	0
71	0
72	0
73	0
74	0
75	0
76	0
77	0
78	0
79	0.0281524905
80	0
81	0
82	0
83	0
84	0
85	0
86	0
87	0
88	0
89	0.0240237292
90	0
91	0.0256989629
92	0
93	0
94	0
95	0.0241214181
96	0.0255371655
97	0
98	0.0264997814
99	0.0278952319
100	0
101	0.0295693105
102	0
103	0
104	0
105	0.0292830917
106	0.0306799449
107	0
108	0
109	0
110	0
111	0
112	0
113	0
114	0
115	0
116	0
117	0
118	0
119	0
120	0.0266439221
121	0.02866478
122	0.0301052797
123	0
124	0.0310747436
125	0

};
%
%
%

\end{axis}
\end{tikzpicture}}
  \\
  \vspace{2pt}
  \caption{Normalized average LLR magnitude at each non-frozen bit index for $PC(1024,170)$ at $E_b/N_0 = 2.5$ dB and $C=7$, ${\rm FER_t}=10^{-4}$. Indices highlighted in red correspond to the non-zero $f^{E_1}$ shown in Fig. \ref{fig:spikes}.}
  \label{fig:avgLLRs}
\end{figure}
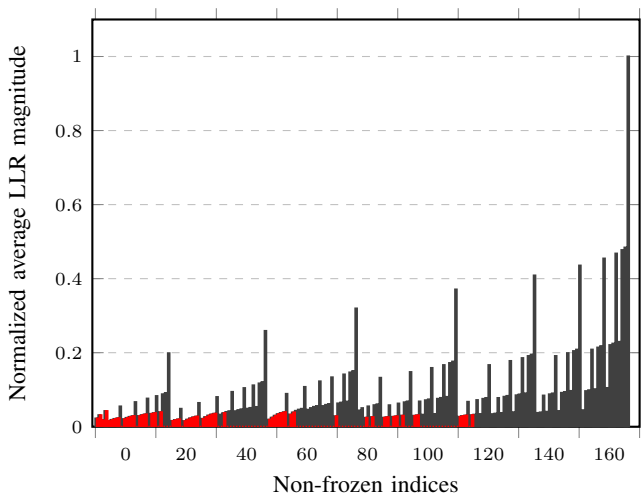

Once the critical set $\boldsymbol{\mathcal{I}^{C}}$ has been identified, the trend observed in Fig. \ref{fig:avgLLRs-TIS} can be used to apply further refinement to the bit-flipping search space through an LLR threshold. Critical indices are considered to be possibly in error only if their LLR is lower than the threshold. We call this technique thresholded index selection (TIS), while SC-Flip incorporating TIS is identified as thresholded SC-Flip (TSCF). After TIS, the identified indices are flipped starting from the leftmost one. This is due to the fact that LLR magnitudes are affected by the propagation of errors, and a small LLR can be caused by a previous channel-induced wrong decision. Thus, selecting the $T_{max}$ indices with the lowest LLR magnitude, like in the original SC-Flip \cite{SCFlip14}, would include critical indices that have a lower $f^{E_1}$, but have been adversely affected by the earlier wrong decision. This approach gives superior error-correction performance with respect to the LLR-based sorting, and substitutes the sorter with a low-complexity threshold comparator.

\subsection{LLR Threshold Selection}

The effectiveness of TIS and TSCF is based on the significant difference in magnitude between the average LLR values of the critical indices in case of successful and failed decoding, that allows the identification of an LLR threshold. Let us identify this threshold as $\Omega$. In order to maximize the accuracy of the thresholding process, the distribution of LLR values for critical indices needs to be observed.

\begin{figure}
  \centering
   \scalebox{1}{\begin{tikzpicture}
  \pgfplotsset{
    label style = {font=\fontsize{9pt}{7.2}\selectfont},
    tick label style = {font=\fontsize{7pt}{7.2}\selectfont}
  }

\begin{axis}[
	scale = 1,
    ylabel style={align=center},
    xlabel={Normalized LLR Magnitude}, xlabel style={yshift=0.4em},
    ylabel={Normalized Frequency}, ylabel style={yshift=-0.75em},
    grid=both,
    ymajorgrids=true,
    xmajorgrids=true,
    grid style=dashed,
    width=\columnwidth, height=7cm,
    thick,
    mark size=3,
    legend style={
      anchor={center},
      cells={anchor=west},
      column sep= 2mm,
      font=\fontsize{7pt}{7.2}\selectfont,
    },
    legend to name=errorhistogram,
    legend columns=2,
]

\addplot[
    color=blue,
    mark=square,
    thick,
    mark size=3,
]
table {
0.0333340128	0.0785350859
0.0666673061	0.1346618614
0.1000005995	0.2180623278
0.1333338929	0.3238075048
0.1666676658	0.4584799661
0.2000002398	0.6057133771
0.2333340128	0.7557981768
0.2666677857	0.8798918804
0.3000003597	0.9678715285
0.3333341327	1
0.3666667066	0.9697053212
0.4000004796	0.884248463
0.4333342526	0.7609285563
0.4666668265	0.6128365487
0.5000005995	0.4671825313
0.5333331735	0.3328068688
0.5666669464	0.2243269027
0.6000007194	0.1403328387
0.6333332934	0.0838138647
0.6666670663	0.0457706169
0.7000008393	0.0250900996
0.7333334133	0.0128471486
0.7666671862	0.0057769769
0.7999997602	0.0025227899
0.8333335332	0.0011977952
0.8666673061	0.000498198
0.8999998801	0.0001695993
0.9333336531	3.17998728005088E-05
0.966666227	3.17998728005088E-05
1	3.17998728005088E-05
};
\addlegendentry{Correct bit}

\addplot[
    color=red,
    mark=o,
    thick,
    mark size=3,
]
table {
0.0348453692	1
0.0696875011	0.5050202948
0.1045297529	0.2377697073
0.1393721246	0.1042512284
0.1742142565	0.0448622089
0.2090563884	0.0164494766
0.2438985203	0.0057679983
0.2787406522	0.001495407
};
\addlegendentry{$E_1$}

\addplot[
    color=black,
    thick,
    mark size=3,
]
table {
0.1017310014  1.0
0.1017310014 -0.0
};


\end{axis}
\end{tikzpicture}}
   \\
   \ref{errorhistogram}
  \caption{Distribution of LLR magnitude in case of correct bit estimation (blue) and $E_1$ (red) for critical index $\mathcal{I}^C_{0}$ for $PC(1024,170)$ and $C=7$ at $E_b/N_0 = 3.25$ dB.}
  \label{fig:errorhistogram}
\end{figure}
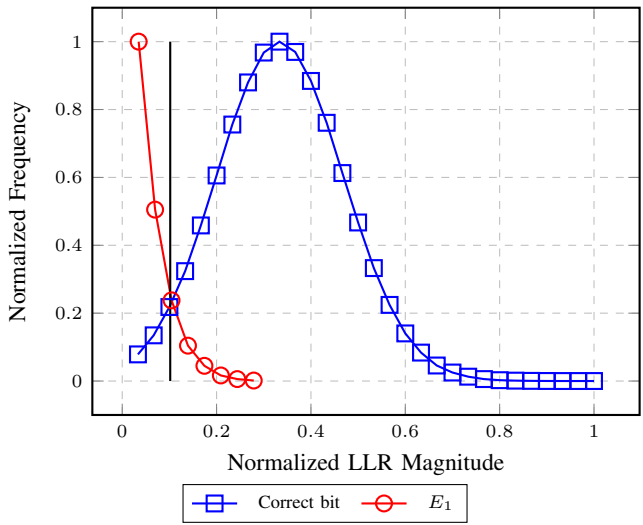

While Fig.~\ref{fig:avgLLRs-TIS} depicts the LLR magnitude gap between the average LLR magnitudes for $E_1$ occurrence and correct bit estimation, Fig.~\ref{fig:errorhistogram} shows the distributions of LLR magnitudes for the first index in the critical set $\mathcal{I}^{C}_{0}$ in case $E_1$ occurred there (red line) and when it did not (blue line). The vertical line in Fig.~\ref{fig:errorhistogram} shows where the two distributions overlap, and represents an ideal thresholding point. In fact, it is very likely that LLRs to the left of the line represent erroneous decisions, while the LLRs to its right are very likely to represent a correct estimation. 

A possible approach to thresholding is to set a different $\Omega$ for each index at the crossover point between the correct and erroneous LLR distributions. However, this approach results in large memory consumption and the threshold list depends on the size of $\boldsymbol{\mathcal{I}^C}$. It can however be observed in Fig.~\ref{fig:avgLLRs-TIS} that while the mean of correct LLRs varies with each critical index, the mean of erroneous LLRs remains very low. Consequently, for a target $E_b/N_0$, a single threshold for entire set $\boldsymbol{\mathcal{I}}^C$ can be selected. In Section \ref{sec:simresults}, it is shown that TSCF with a single threshold for the whole critical set is sufficient to approach the SCO-1 performance within a small number of iterations.

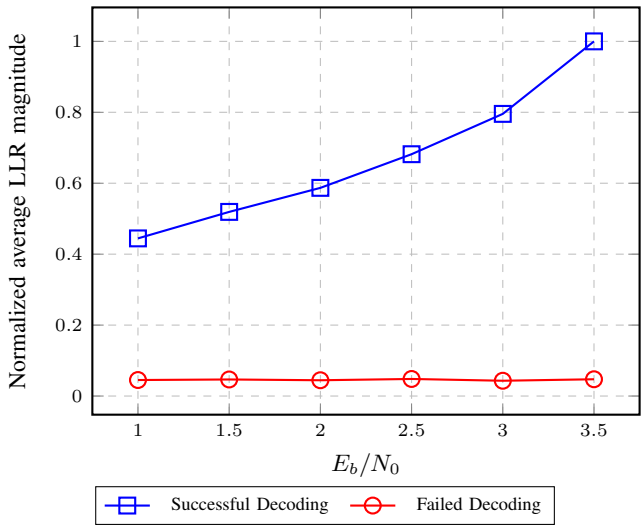
\begin{figure}
  \centering
   \scalebox{1}{\begin{tikzpicture}
  \pgfplotsset{
    label style = {font=\fontsize{9pt}{7.2}\selectfont},
    tick label style = {font=\fontsize{7pt}{7.2}\selectfont}
  }

\begin{axis}[
	scale = 1,
    ylabel style={align=center},
    xlabel={$E_b/N_0$}, xlabel style={yshift=0.4em},
    ylabel={Normalized average LLR magnitude}, ylabel style={yshift=-0.75em},
    grid=both,
    ymajorgrids=true,
    xmajorgrids=true,
    grid style=dashed,
    width=\columnwidth, height=7cm,
    thick,
    mark size=3,
    legend style={
      anchor={center},
      cells={anchor=west},
      column sep= 2mm,
      font=\fontsize{7pt}{7.2}\selectfont,
    },
    legend to name=sortederrors,
    legend columns=2,
]

\addplot[
    color=blue,
    mark=square,
    thick,
    mark size=3,
]
table {
1.0	0.4444842869
1.5	0.5191646511
2.0	0.5869306819
2.5	0.681842746
3.0	0.7955091459
3.5	1
};
\addlegendentry{Successful Decoding}

\addplot[
    color=red,
    mark=o,
    thick,
    mark size=3,
]
table {
1.0	0.0451345558
1.5	0.0466919894
2.0	0.0444987589
2.5	0.0481177634
3.0	0.0431048513
3.5	0.047294925
};
\addlegendentry{Failed Decoding}

\end{axis}
\end{tikzpicture}}
   \\
   \ref{sortederrors}
  \caption{Normalized LLR magnitude averaged over the whole $\boldsymbol{\mathcal{I}}^C$ with respect to $E_b/N_0$, in case of successful (blue) and failed (red) decoding;~ $PC(1024,170)$ and $C=7$.}
  \label{fig:avLLRvsSNR}
\end{figure}

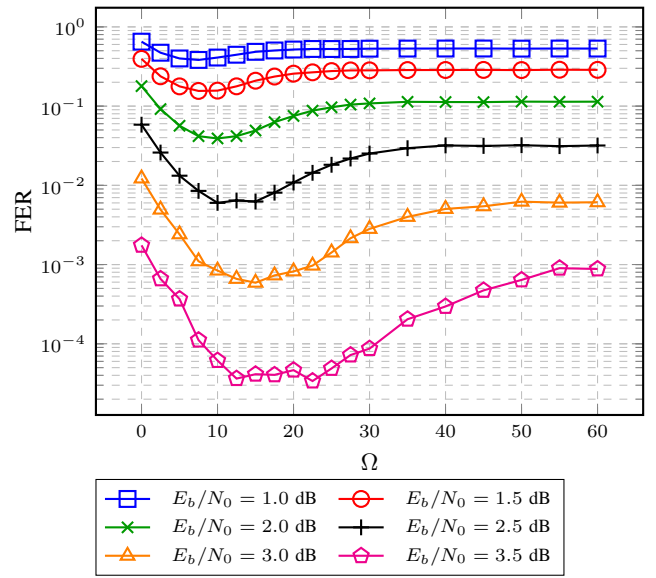
\begin{figure}
  \centering
   \scalebox{1}{\begin{tikzpicture}
  \pgfplotsset{
    label style = {font=\fontsize{9pt}{7.2}\selectfont},
    tick label style = {font=\fontsize{7pt}{7.2}\selectfont}
  }

\begin{axis}[
	scale = 1,
    ymode=log,
    xlabel={$\Omega$}, xlabel style={yshift=0.4em},
    ylabel={FER}, ylabel style={yshift=-0.75em},
    grid=both,
    ymajorgrids=true,
    xmajorgrids=true,
    grid style=dashed,
    width=\columnwidth, height=7cm,
    thick,
    mark size=3,
    legend style={
      anchor={center},
      cells={anchor=west},
      column sep= 2mm,
      font=\fontsize{7pt}{7.2}\selectfont,
    },
    legend to name=LLRTHRvsFER_R016,
    legend columns=2,
]

\addplot[
    color=blue,
    mark=square,
    thick,
    mark size=3,
]
table {
0	6.50870e-01 
2.5	4.72100e-01 
5	3.99610e-01 
7.5	3.82790e-01 
10	4.10220e-01 
12.5	4.45270e-01 
15	4.84520e-01 
17.5	5.03790e-01 
20	5.17920e-01 
22.5	5.26390e-01 
25	5.25770e-01 
27.5	5.26510e-01 
30	5.30660e-01
35	5.32310e-01 
40	5.33080e-01 
45	5.33380e-01 
50	5.31440e-01 
55	5.32880e-01 
60	5.32420e-01
};
\addlegendentry{$E_b/N_0 = 1.0$ dB}

\addplot[
    color=red,
    mark=o,
    thick,
    mark size=3,
]
table {
0	3.94800e-01 
2.5	2.37910e-01 
5	1.77530e-01 
7.5	1.55440e-01 
10	1.56860e-01 
12.5	1.77720e-01 
15	2.09240e-01 
17.5	2.36010e-01 
20	2.56830e-01 
22.5	2.66930e-01 
25	2.76850e-01 
27.5	2.82310e-01 
30	2.82720e-01
35	2.85270e-01 
40	2.85670e-01 
45	2.86310e-01  
50	2.85070e-01 
55	2.87940e-01 
60	2.86880e-01
};
\addlegendentry{$E_b/N_0 = 1.5$ dB}

\addplot[
    color=green!60!black,
    mark=x,
    thick,
    mark size=3,
]
table {
0	1.78760e-01 
2.5	9.17400e-02 
5	5.65100e-02 
7.5	4.17400e-02 
10	3.90900e-02 
12.5	4.17900e-02 
15	4.92200e-02 
17.5	6.28000e-02 
20	7.51500e-02 
22.5	8.76800e-02 
25	9.66300e-02 
27.5	1.04610e-01 
30	1.08180e-01
35	1.13280e-01 
40	1.12620e-01 
45	1.12310e-01
50	1.13940e-01 
55	1.13470e-01 
60	1.13800e-01 

};
\addlegendentry{$E_b/N_0 = 2.0$ dB}

\addplot[
    color=black,
    mark=+,
    thick,
    mark size=3,
]
table {
0	5.82900e-02 
2.5	2.59300e-02 
5	1.32300e-02 
7.5	8.51000e-03 
10	6.01000e-03 
12.5	6.44000e-03 
15	6.25000e-03 
17.5	8.08000e-03 
20	1.08500e-02 
22.5	1.43900e-02 
25	1.81800e-02 
27.5	2.18000e-02 
30	2.52100e-02 
35	2.94300e-02 
40	3.18800e-02 
45	3.14200e-02 
50	3.21000e-02 
55	3.12400e-02 
60	3.18300e-02 
};
\addlegendentry{$E_b/N_0 = 2.5$ dB}

\addplot[
    color=orange,
    mark=triangle,
    thick,
    mark size=3,
]
table {
0	1.22100e-02 
2.5	4.94000e-03 
5	2.41000e-03 
7.5	1.10000e-03 
10	8.40000e-04 
12.5	6.60000e-04 
15	5.90000e-04 
17.5	7.30000e-04 
20	8.20000e-04 
22.5	9.70000e-04 
25	1.42000e-03 
27.5	2.15000e-03 
30	2.82000e-03
35	4.00000e-03 
40	5.04000e-03 
45	5.44000e-03 
50	6.22000e-03 
55	6.05000e-03 
60	6.12000e-03
};
\addlegendentry{$E_b/N_0 = 3.0$ dB}

\addplot[
    color=magenta,
    mark=pentagon,
    thick,
    mark size=3,
]
table {
0	1.75000e-03 
2.5	6.60000e-04 
5	3.67531e-04 
7.5	1.11277e-04 
10	6.14778e-05 
12.5	3.63750e-05 
15	4.13178e-05 
17.5	4.05539e-05 
20	4.64944e-05 
22.5	3.38768e-05 
25	4.88388e-05 
27.5	7.21314e-05 
30	8.69766e-05
35	2.04186e-04 
40	2.96021e-04 
45	4.72438e-04 
50	6.40000e-04 
55	9.00000e-04 
60	8.80000e-04 
};
\addlegendentry{$E_b/N_0 = 3.5$ dB}

\end{axis}
\end{tikzpicture}}
   \\
   \ref{LLRTHRvsFER_R016}
  \caption{FER curves for TSCF decoding with various $\Omega$ values for $PC(1024,170)$, $C=7$, $T_{max}=10$.}
  \label{fig:LLRTHRvsFER_R016}
\end{figure}

Average LLR magnitudes do not only depend on the critical index, but also on $E_b/N_0$. A similar trend is in fact observed in Fig.~\ref{fig:avLLRvsSNR}, where it is shown that while the average LLRs for critical indices tend to increase with $E_b/N_0$ in case of correct estimations, the erroneous LLR average remains around the same magnitude, resulting in an increasing gap between them. As a result, there is a different threshold value for each $E_b/N_0$ that maximizes the error-correction performance gain of TSCF. Based on the observations from Fig.~\ref{fig:avgLLRs-TIS}, Fig.~\ref{fig:errorhistogram} and Fig.~\ref{fig:avLLRvsSNR}, an LLR threshold sweep is performed for TSCF decoding. Fig.~\ref{fig:LLRTHRvsFER_R016} presents the FER of $PC(1024,170)$ for various $E_b/N_0$ values for a set of $\Omega$ values, with $C=7$ and $T_{max}=10$. The critical set $\boldsymbol{\mathcal{I}^C}$ is obtained at $E_b/N_0 =3.0$ dB and includes $56$ indices. The best FER can be obtained by selecting a different threshold value for each $E_b/N_0$. However, if a single LLR threshold for all $E_b/N_0$ points is more suitable for the application, it should be selected in accordance with the target FER. It can be seen that the $\Omega$ value that corresponds to best FER value tends to increase as the channel conditions improve. Also note that, with $\Omega=0$, the LLR values of critical indices will always fall above the threshold, and TSCF reverts to SC decoding with $K+C$ non-frozen bits. 

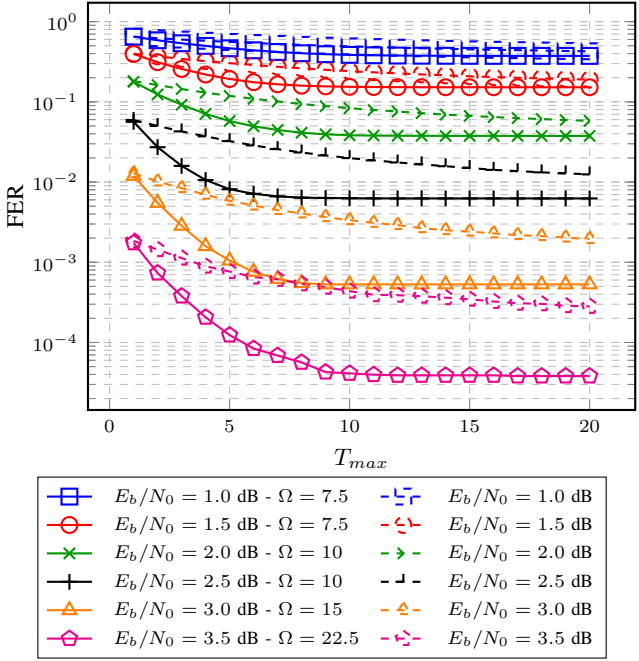
\begin{figure}
  \centering
   \scalebox{1}{\begin{tikzpicture}
  \pgfplotsset{
    label style = {font=\fontsize{9pt}{7.2}\selectfont},
    tick label style = {font=\fontsize{7pt}{7.2}\selectfont}
  }

\begin{axis}[
	scale = 1,
    ymode=log,
    xlabel={$T_{max}$}, xlabel style={yshift=0.4em},
    ylabel={FER}, ylabel style={yshift=-0.75em},
    grid=both,
    ymajorgrids=true,
    xmajorgrids=true,
    grid style=dashed,
    width=\columnwidth, height=7cm,
    thick,
    mark size=3,
    legend style={
      anchor={center},
      cells={anchor=west},
      column sep= 2mm,
      font=\fontsize{7pt}{7.2}\selectfont,
    },
    legend to name=TMAXvsFER_R016,
    legend columns=2,
]

\addplot[
    color=blue,
    mark=square,
    thick,
    mark size=3,
]
table {
1	0.64952
2	0.58541
3	0.53734
4	0.49848
5	0.4653
6	0.4389
7	0.41839
8	0.40341
9	0.39261
10	0.38518
11	0.38061
12	0.37783
13	0.37634
14	0.37527
15	0.37475
16	0.37432
17	0.37411
18	0.37406
19	0.37402
20	0.37401
};
\addlegendentry{$E_b/N_0 = 1.0$ dB - $\Omega=7.5$}

\addplot[
    color=blue,
    dashed,
    mark=square,
    thick,
    mark size=3,
]
table {
1	0.64954
2	0.62158
3	0.5972
4	0.57677
5	0.55825
6	0.54187
7	0.52706
8	0.51376
9	0.50186
10	0.49094
11	0.48143
12	0.47289
13	0.46509
14	0.45802
15	0.45147
16	0.44576
17	0.44075
18	0.43589
19	0.43113
20	0.42688
};
\addlegendentry{$E_b/N_0 = 1.0$ dB}

\addplot[
    color=red,
    mark=o,
    thick,
    mark size=3,
]
table {
1	0.39514
2	0.31198
3	0.257
4	0.2191
5	0.19261
6	0.17564
7	0.16539
8	0.15915
9	0.15587
10	0.15399
11	0.15294
12	0.1525
13	0.15219
14	0.1521
15	0.15205
16	0.15201
17	0.15201
18	0.152
19	0.152
20	0.152
};
\addlegendentry{$E_b/N_0 = 1.5$ dB - $\Omega=7.5$}

\addplot[
    color=red,
    dashed,
    mark=o,
    thick,
    mark size=3,
]
table {
1	0.39822
2	0.36851
3	0.34303
4	0.3216
5	0.30366
6	0.28781
7	0.27362
8	0.26149
9	0.25071
10	0.24115
11	0.23347
12	0.22647
13	0.21977
14	0.21368
15	0.20837
16	0.2038
17	0.19967
18	0.19595
19	0.19263
20	0.18956
};
\addlegendentry{$E_b/N_0 = 1.5$ dB}

\addplot[
    color=green!60!black,
    mark=x,
    thick,
    mark size=3,
]
table {
1	0.17764
2	0.12378
3	0.092
4	0.07089
5	0.05762
6	0.04959
7	0.04461
8	0.04123
9	0.03937
10	0.03833
11	0.03797
12	0.03781
13	0.03772
14	0.03765
15	0.03765
16	0.03764
17	0.03763
18	0.03761
19	0.03761
20	0.03761
};
\addlegendentry{$E_b/N_0 = 2.0$ dB - $\Omega=10$}

\addplot[
    color=green!60!black,
    dashed,
    mark=x,
    thick,
    mark size=3,
]
table {
1	0.1815
2	0.16177
3	0.14453
4	0.13007
5	0.11872
6	0.10919
7	0.10122
8	0.09397
9	0.0884
10	0.08342
11	0.07905
12	0.07569
13	0.07233
14	0.06935
15	0.06687
16	0.06463
17	0.06271
18	0.061
19	0.05928
20	0.05789
};
\addlegendentry{$E_b/N_0 = 2.0$ dB}

\addplot[
    color=black,
    mark=+,
    thick,
    mark size=3,
]
table {
1	0.05638
2	0.02719
3	0.01592
4	0.0106
5	0.00814
6	0.00714
7	0.00663
8	0.00639
9	0.00633
10	0.00628
11	0.00626
12	0.00625
13	0.00625
14	0.00625
15	0.00625
16	0.00625
17	0.00625
18	0.00625
19	0.00625
20	0.00625
};
\addlegendentry{$E_b/N_0 = 2.5$ dB - $\Omega=10$}

\addplot[
    color=black,
    dashed,
    mark=+,
    thick,
    mark size=3,
]
table {
1	0.05941
2	0.05006
3	0.04283
4	0.03715
5	0.03241
6	0.02863
7	0.02575
8	0.02336
9	0.02149
10	0.0199
11	0.01865
12	0.01755
13	0.01665
14	0.01575
15	0.01499
16	0.01419
17	0.0137
18	0.01321
19	0.01285
20	0.01246
};
\addlegendentry{$E_b/N_0 = 2.5$ dB}

\addplot[
    color=orange,
    mark=triangle,
    thick,
    mark size=3,
]
table {
1	0.01192
2	0.0055
3	0.00285
4	0.00159
5	0.00105
6	0.00076
7	0.00062
8	0.00055
9	0.00053
10	0.00053
11	0.00053
12	0.00053
13	0.00053
14	0.00053
15	0.00053
16	0.00053
17	0.00053
18	0.00053
19	0.00053
20	0.00053
};
\addlegendentry{$E_b/N_0 = 3.0$ dB - $\Omega=15$}

\addplot[
    color=orange,
    dashed,
    mark=triangle,
    thick,
    mark size=3,
]
table {
1	0.01286
2	0.01016
3	0.0084
4	0.00695
5	0.00584
6	0.00509
7	0.00448
8	0.00406
9	0.00365
10	0.00335
11	0.0031
12	0.00291
13	0.00266
14	0.00251
15	0.00241
16	0.00229
17	0.00218
18	0.00212
19	0.00203
20	0.00195
};
\addlegendentry{$E_b/N_0 = 3.0$ dB}

\addplot[
    color=magenta,
    mark=pentagon,
    thick,
    mark size=3,
]
table {
1	0.0017442748
2	0.0007343511
3	0.000378626
4	0.0002061069
5	0.0001236641
6	8.39694656488549E-05
7	6.87022900763359E-05
8	5.64885496183206E-05
9	4.27480916030534E-05
10	4.12213740458015E-05
11	3.96946564885496E-05
12	3.89312977099237E-05
13	3.89312977099237E-05
14	3.89312977099237E-05
15	3.89312977099237E-05
16	3.89312977099237E-05
17	3.81679389312977E-05
18	3.81679389312977E-05
19	3.81679389312977E-05
20	3.81679389312977E-05
};
\addlegendentry{$E_b/N_0 = 3.5$ dB - $\Omega=22.5$}

\addplot[
    color=magenta,
    dashed,
    mark=pentagon,
    thick,
    mark size=3,
]
table {
1	0.0018457447
2	0.0014361702
3	0.0011276596
4	0.000893617
5	0.0007659574
6	0.0006542553
7	0.0006117021
8	0.000537234
9	0.0004893617
10	0.0004361702
11	0.000393617
12	0.0003882979
13	0.0003776596
14	0.0003617021
15	0.0003457447
16	0.0003244681
17	0.0003085106
18	0.0003031915
19	0.000287234
20	0.0002819149
};
\addlegendentry{$E_b/N_0 = 3.5$ dB}

\end{axis}
\end{tikzpicture}}
   \\
   \ref{TMAXvsFER_R016}
  \caption{FER curves for TSCF decoding (solid lines) and SC-Flip (dashed lines) with various $T_{max}$ values, for $PC(1024,170)$, $C=7$.}
  \label{fig:TMAXvsFER_R016}
\end{figure}

Based on the information obtained on the impact of $\Omega$ on FER, TSCF has been compared to the baseline SC-Flip algorithm. Fig. \ref{fig:TMAXvsFER_R016} shows the change in FER with respect to $T_{max}$ for both algorithms at various $E_b/N_0$ values. The same critical set used in Fig.~\ref{fig:LLRTHRvsFER_R016} is used here. Solid curves represent TSCF with the optimal $\Omega$ value for each $E_b/N_0$, and dashed lines represent SC-Flip. Simulations are performed for $PC(1024,170)$ for up to maximum $T_{max}$ of $20$. It can be seen that while the error-correction performance of SC-Flip improves almost linearly with $T_{max}$, TSCF quickly converges to a lower bound in each case, i.e. the SCO-1 performance. This is due to the fact that TSCF is able to find the correct critical index quicker than SC-Flip. As a result, TSCF can reach a target FER with a smaller $T_{max}$ value, which results in a shorter worst case decoding latency and more stable average latency. For example, it takes only $5$ iterations for TSCF to achieve FER $=10^{-3}$ at $E_b/N_0 = 3.0$ dB, while it takes $20$ iterations for SC-Flip. From another point of view, given a fixed maximum number of iterations $T_{max} > 1$, TSCF results in lower FER than SC-Flip in all cases. For example. with $T_{max} = 5$, TSCF has FER=$1.23 \times 10^{-4}$ at $E_b/N_0 = 3.5$ dB, where it is $7.65 \times 10^{-4}$ for SC-Flip. Finally, note the crossing of the SC-Flip curve at $E_b/N_0 = 3.5$ dB with TSCF at $E_b/N_0 = 3.0$ dB. This crossover means that, at $T_{max} \in \{6,7,8\}$, both SC-Flip and TSCF have identical error-correction performance; however TSCF achieves this while having an $E_b/N_0$ value that is $0.5$ dB less than that of SC-Flip.

\subsection{Critical Set Selection}\label{sec:indexselection}

The identification of $\boldsymbol{\mathcal{I}^{C}}$ using SC-Oracle helps improve error-correction performance and reduce $T_{max}$. However, not every critical index in Fig.~\ref{fig:spikes} has the same $f^{E_1}$. In fact given ${\rm FER_t}$, at each $E_b/N_0$ SC-Oracle reports $\boldsymbol{\mathcal{I}^{C}}$ of different sizes and composed of different indices. For example, in Fig.~\ref{fig:spikes}, although the $6$ indices with the largest $f^{E_1}$ are the same for considered $E_b/N_0$ points, the index with $7^{\rm th}$ largest $f^{E_1}$ is different. It should be also noted that SC-Oracle identifies a smaller list of indices as $E_b/N_0$ grows, which is a subset of that identified for lower $E_b/N_0$ values.

It can be seen in Fig.~\ref{fig:spikes} that the majority of $f^{E_1}$ is concentrated around a small set of non-frozen indices. For example, the first seven indices that are most likely to incur an error account for $75.3\%$ of $f^{E_1}$. As we addressed in Section \ref{sec:ourtech}, it is thus unnecessary to consider all indices that have nonzero $f^{E_1}$ as part of the critical set, especially considering the limits imposed by $T_{max}$. A large critical set, together with a small $T_{max}$, can lead to substantial error-correction performance degradation. Moreover, in order to minimize the TSCF decoder implementation cost, a single critical set is desirable. A first critical set selection approach is to use the $\boldsymbol{\mathcal{I}^{C}}$ derived from the lowest considered $E_b/N_0$, given the fact that at higher $E_b/N_0$ the critical set is a subset of this one. With a high enough $T_{max}$, such a set can correct all $E_1$ at higher $E_b/N_0$ values, achieving the SCO-1 performance. On the other hand, some of the critical indices only occur at low $E_b/N_0$ values, introducing unnecessary latency and memory usage for iterations at high $E_b/N_0$ points. Thus, a more efficient approach selects the set of indices starting from the full set identified for a high $E_b/N_0$ value. For example, out of the $56$ critical indices identified after omitting the ones with low $f^{E_1}$ for $PC(1024,170)$ with $C=7$ at $E_b/N_0 = 3.0$ dB, all of them exist within the $70$ indices found at $E_b/N_0 = 2.5$ dB, and they are within the most critical ones. Similarly, they in turn constitute a subset of the $81$ indices found at $E_b/N_0 = 2.0$ dB. Finally, where all the indices of $E_b/N_0 = 3.0$ dB represent $100\%$ of all $E_1$ occurrences, the same set of indices covers $99.99\%$ and $99.96\%$ of all $E_1$ errors for when $E_b/N_0 = 2.5$ and $2.0$ dB, respectively. As a result, a single $\boldsymbol{\mathcal{I}}^C$ identified at a high $E_b/N_0$ value can be selected with negligible degradation in error-correction performance at lower $E_b/N_0$ values.

It should be noted that the discussed critical index selection scenarios can also be applied to the FIS and EIS criteria described in \cite{SCF-WCNC18}. Finally, the indices and size of the critical set may change with respect to polar code construction.

\subsection{TSCF Decoding for Higher Rate Polar Codes}\label{sec:highrate}

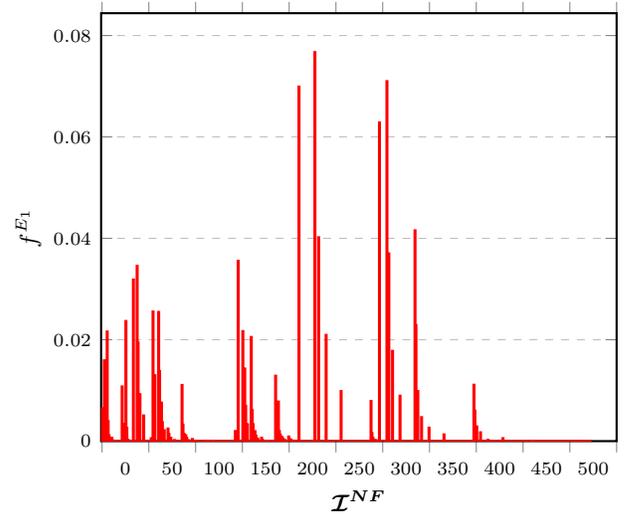
\begin{figure}
  \centering
   \scalebox{1}{\begin{tikzpicture}
  \pgfplotsset{
    label style = {font=\fontsize{9pt}{7.2}\selectfont},
    tick label style = {font=\fontsize{7pt}{7.2}\selectfont},
    yticklabel style={/pgf/number format/fixed},
    scaled y ticks=false
  }

\begin{axis}[
	scale = 1,
    ybar interval,
    xlabel={$\boldsymbol{\mathcal{I}^{NF}}$}, xlabel style={yshift=0em},
    ylabel={$f^{E_1}$}, ylabel style={yshift=-0.75em},
    grid=both,
    xmin = -1,
    xmax = 550,
    ymin = 0,
    xtick={0,50,100,150,200,250,300,350,400,450,500,550},
    ymajorgrids=true,  
    xmajorgrids=false,
    grid style=dashed,
    thick,
    mark size=3,
]

\addplot[
    bar width=5pt,
    red,
    thick,
]
table {
0	0.006432336
1	8.29978835539694E-05
2	0.0159355936
3	0.007096319
4	0
5	0.0216209487
6	0.0039423995
7	0.0011204714
8	0.0003734905
9	0
10	0.0006639831
11	0.0001244968
12	8.29978835539694E-05
13	0
14	4.14989417769847E-05
15	0
16	0
17	0
18	0
19	0
20	0
21	0.0107897249
22	0.0033614143
23	0.0018259534
24	0
25	0.0236958958
26	0.0026559323
27	0.0002904926
28	0.0001244968
29	8.29978835539694E-05
30	0
31	0
32	0
33	0.0318711873
34	0
35	0
36	0
37	0.0345686185
38	0.0194215048
39	0
40	0.009254264
41	0
42	0
43	0
44	0.0050628709
45	0
46	0
47	0
48	0
49	0
50	0
51	0
52	0.0005394862
53	4.14989417769847E-05
54	0.0255633481
55	0
56	0.0129891688
57	0
58	0
59	0
60	0.0254803503
61	0.0137776487
62	0
63	0.0075943063
64	0.0037349048
65	0
66	0.002157945
67	0
68	0
69	0
70	0.0024899365
71	0.0013694651
72	0
73	0.0006639831
74	0
75	0
76	0
77	0.0002074947
78	0
79	0
80	0
81	0
82	0
83	0
84	0
85	0.0110802175
86	0.0032369175
87	0.001452463
88	0
89	0.0011619704
90	0.0006639831
91	0
92	0.0001659958
93	0
94	0
95	0
96	0.0004149894
97	8.29978835539694E-05
98	0
99	8.29978835539694E-05
100	0
101	0
102	0
103	0
104	0
105	0
106	0
107	0
108	0
109	0
110	0
111	4.14989417769847E-05
112	4.14989417769847E-05
113	0
114	0
115	0
116	0
117	0
118	4.14989417769847E-05
119	0
120	0
121	0
122	0
123	0
124	0
125	0
126	0
127	0
128	0
129	0
130	0
131	0
132	0
133	0
134	0
135	0
136	0
137	0
138	0
139	0
140	0
141	0
142	0.0019919492
143	0.0001244968
144	0
145	0.0355645931
146	0
147	0
148	0
149	0
150	0.0217039465
151	0
152	0.0143171349
153	0.0069303233
154	0
155	0.0033199153
156	0
157	0
158	0
159	0.0205419762
160	0.0061003444
161	0.0033199153
162	0
163	0.0019504503
164	0.0010789725
165	0
166	0.0004564884
167	0
168	0
169	0
170	0.0006639831
171	0.0002074947
172	0
173	8.29978835539694E-05
174	0
175	0
176	0
177	8.29978835539694E-05
178	0
179	0
180	0
181	0
182	0
183	0
184	0
185	0.0129061709
186	0.0068473254
187	0
188	0.0078018011
189	0.0019919492
190	0.0012864672
191	0
192	0.0008299788
193	0.0004564884
194	0
195	0.0002074947
196	0
197	0
198	0
199	0.0009129767
200	0.0003319915
201	0.0003319915
202	0
203	0.0001244968
204	8.29978835539694E-05
205	0
206	0
207	0
208	0
209	0
210	0.0699257169
211	0
212	0
213	0
214	0
215	0
216	0
217	0
218	0
219	0
220	0
221	0
222	0
223	0
224	0
225	0
226	4.14989417769847E-05
227	0.0767315433
228	0
229	0
230	0
231	0.0402124746
232	0
233	0
234	0
235	0
236	0
237	0
238	0
239	0.0209569656
240	0
241	0
242	0
243	0
244	0
245	0
246	0
247	0
248	0
249	0
250	0
251	0
252	0
253	0
254	0
255	0.0098767481
256	0
257	0
258	0
259	0
260	0
261	0
262	0
263	0
264	0
265	0
266	0
267	0
268	0
269	0
270	0
271	0
272	0
273	0
274	0
275	0
276	0
277	0
278	0
279	0
280	0
281	0
282	0
283	0
284	0
285	0
286	0
287	0.0079262979
288	0.0015354608
289	0.0006639831
290	0.0002074947
291	0
292	0.0001659958
293	0.0001244968
294	8.29978835539694E-05
295	0
296	0.0628708968
297	0
298	0
299	0
300	0
301	0
302	0
303	0
304	0.0710046894
305	0
306	0.0370170561
307	0
308	0
309	0
310	0.017803046
311	0
312	0
313	0
314	0
315	0
316	0
317	0
318	0.0089637714
319	0
320	0
321	0
322	0
323	0
324	0
325	0
326	0
327	0
328	0
329	0
330	0
331	0
332	0
333	0
334	0.0415819397
335	0.0228659169
336	0
337	0.0098767481
338	0
339	0
340	0
341	0.0047308794
342	0
343	0
344	0
345	0
346	0
347	0
348	0
349	0.0026559323
350	0
351	0
352	0
353	0
354	0
355	0
356	0
357	0
358	0
359	0
360	0
361	0
362	0
363	0
364	0
365	0.0013279661
366	0
367	0
368	0
369	0
370	0
371	0
372	0
373	0
374	0
375	0
376	0
377	0
378	0
379	0
380	0
381	0
382	0
383	0
384	0
385	0
386	0
387	0
388	0
389	0
390	0
391	0
392	0
393	0
394	0
395	0
396	0
397	0.0111217164
398	0.0060173466
399	0
400	0.0029049259
401	0
402	0
403	0
404	0.0017429556
405	0
406	0
407	0
408	0
409	0
410	0
411	0
412	0.0002489937
413	0
414	0
415	0
416	0
417	0
418	0
419	0
420	0
421	0
422	0
423	0
424	0
425	0
426	0
427	0
428	0.0005809852
429	0
430	0
431	0
432	0
433	0
434	0
435	0
436	0
437	0
438	0
439	0
440	0
441	0
442	0
443	0
444	0
445	0
446	0
447	0
448	0
449	0
450	0
451	0
452	0
453	0
454	0
455	0
456	0
457	0
458	0
459	0
460	8.29978835539694E-05
461	0
462	0
463	0
464	0
465	0
466	0
467	0
468	0
469	0
470	0
471	0
472	0
473	0
474	0
475	0
476	0
477	0
478	0
479	0
480	0
481	0
482	0
483	0
484	0
485	0
486	0
487	0
488	0
489	0
490	0
491	0
492	0
493	0
494	0
495	0
496	0
497	0
498	0
499	0
500	0
501	0
502	0
503	0
504	0
505	0
506	0
507	0
508	0
509	0
510	0
511	0
512	0
513	0
514	0
515	0
516	0
517	0
518	0
519	0
520	0
521	0
522	0
523	0

};

\end{axis}
\end{tikzpicture}}
   \\
  \caption{$f^{E_1}$ values for non-frozen indices of $PC(1024,512)$, $C=12$, $E_b/N_0 = 3.0$ dB.}
  \label{fig:spikes2}
\end{figure}

\begin{figure}
  \centering
   \scalebox{1}{\begin{tikzpicture}
  \pgfplotsset{
    label style = {font=\fontsize{9pt}{7.2}\selectfont},
    tick label style = {font=\fontsize{7pt}{7.2}\selectfont}
  }

\begin{axis}[
	scale = 1,
    ymode=log,
    xlabel={$T_{max}$}, xlabel style={yshift=0.4em},
    ylabel={FER}, ylabel style={yshift=-0.75em},
    grid=both,
    ymajorgrids=true,
    xmajorgrids=true,
    grid style=dashed,
    width=\columnwidth, height=7cm,
    thick,
    mark size=3,
    legend style={
      anchor={center},
      cells={anchor=west},
      column sep= 2mm,
      font=\fontsize{7pt}{7.2}\selectfont,
    },
    legend to name=TMAXvsFER_R05,
    legend columns=2,
]

\addplot[
    color=blue,
    mark=square,
    thick,
    mark size=3,
]
table {
1	0.81767
2	0.77957
3	0.74969
4	0.72385
5	0.70333
6	0.68539
7	0.67208
8	0.66062
9	0.65185
10	0.64527
11	0.64012
12	0.6362
13	0.63309
14	0.63067
15	0.62915
16	0.62808
17	0.62712
18	0.62664
19	0.62628
20	0.62601
};
\addlegendentry{$E_b/N_0 = 1.0$ dB - $\Omega=5$}

\addplot[
    color=blue,
    dashed,
    mark=square,
    thick,
    mark size=3,
]
table {
1	0.81908
2	0.79292
3	0.77228
4	0.75665
5	0.74369
6	0.73227
7	0.722
8	0.71293
9	0.70499
10	0.69738
11	0.69097
12	0.68536
13	0.68025
14	0.67561
15	0.67105
16	0.66708
17	0.66315
18	0.65988
19	0.65683
20	0.6537
};
\addlegendentry{$E_b/N_0 = 1.0$ dB}

\addplot[
    color=red,
    mark=o,
    thick,
    mark size=3,
]
table {
1	0.44346
2	0.36165
3	0.30864
4	0.27431
5	0.25051
6	0.2354
7	0.2255
8	0.21876
9	0.21506
10	0.21266
11	0.21111
12	0.21027
13	0.20955
14	0.20932
15	0.20919
16	0.20913
17	0.20905
18	0.20901
19	0.20901
20	0.20901
};
\addlegendentry{$E_b/N_0 = 1.5$ dB - $\Omega=5$}

\addplot[
    color=red,
    dashed,
    mark=o,
    thick,
    mark size=3,
]
table {
1	0.43945
2	0.3977
3	0.36898
4	0.34576
5	0.32779
6	0.31323
7	0.30172
8	0.29108
9	0.28168
10	0.27412
11	0.26707
12	0.26076
13	0.25507
14	0.25
15	0.24558
16	0.24106
17	0.23763
18	0.23389
19	0.23059
20	0.22769
};
\addlegendentry{$E_b/N_0 = 1.5$ dB}

\addplot[
    color=green!60!black,
    mark=x,
    thick,
    mark size=3,
]
table {
1	0.12857
2	0.08926
3	0.06576
4	0.05135
5	0.04211
6	0.03581
7	0.03216
8	0.02941
9	0.02769
10	0.0267
11	0.02598
12	0.02547
13	0.02516
14	0.0249
15	0.0248
16	0.02471
17	0.02468
18	0.02465
19	0.02462
20	0.02461
};
\addlegendentry{$E_b/N_0 = 2.0$ dB - $\Omega=7.5$}

\addplot[
    color=green!60!black,
    mark=x,
    dashed,
    thick,
    mark size=3,
]
table {
1	0.12756
2	0.10307
3	0.08828
4	0.07789
5	0.06969
6	0.06338
7	0.05838
8	0.05418
9	0.05095
10	0.04793
11	0.04551
12	0.04339
13	0.04148
14	0.03979
15	0.03837
16	0.03705
17	0.03611
18	0.03503
19	0.03398
20	0.03316
};
\addlegendentry{$E_b/N_0 = 2.0$ dB}

\addplot[
    color=black,
    mark=+,
    thick,
    mark size=3,
]
table {
1	0.0223
2	0.01239
3	0.00764
4	0.00497
5	0.00356
6	0.00275
7	0.00218
8	0.00188
9	0.00174
10	0.00164
11	0.00156
12	0.00151
13	0.00148
14	0.00146
15	0.00146
16	0.00146
17	0.00145
18	0.00144
19	0.00144
20	0.00144
};
\addlegendentry{$E_b/N_0 = 2.5$ dB - $\Omega=10$}

\addplot[
    color=black,
    mark=+,
    dashed,
    thick,
    mark size=3,
]
table {
1	0.02113
2	0.01424
3	0.01101
4	0.00906
5	0.00751
6	0.00641
7	0.00562
8	0.00495
9	0.00451
10	0.00418
11	0.00367
12	0.0034
13	0.00319
14	0.00307
15	0.00287
16	0.00269
17	0.00256
18	0.00245
19	0.00238
20	0.00231
};
\addlegendentry{$E_b/N_0 = 2.5$ dB}

\addplot[
    color=orange,
    mark=triangle,
    thick,
    mark size=3,
]
table {
1	0.0024475806
2	0.0011236559
3	0.0006223118
4	0.000344086
5	0.0001935484
6	0.0001384409
7	0.0001008065
8	0.000077957
9	7.25806451612903E-05
10	7.12365591397849E-05
11	6.98924731182796E-05
12	6.72043010752688E-05
13	6.72043010752688E-05
14	6.72043010752688E-05
15	6.72043010752688E-05
16	6.72043010752688E-05
17	6.72043010752688E-05
18	6.72043010752688E-05
19	6.72043010752688E-05
20	6.72043010752688E-05
};
\addlegendentry{$E_b/N_0 = 3.0$ dB - $\Omega=12.5$}

\addplot[
    color=orange,
    dashed,
    mark=triangle,
    thick,
    mark size=3,
]
table {
1	0.0025185784
2	0.0012859451
3	0.0008626817
4	0.0006090468
5	0.000447496
6	0.000365105
7	0.000279483
8	0.0002455574
9	0.0002116317
10	0.000180937
11	0.0001567044
12	0.0001437803
13	0.0001324717
14	0.0001179321
15	0.0001130856
16	0.0001033926
17	0.000098546
18	9.36995153473344E-05
19	0.000088853
20	8.40064620355412E-05
};
\addlegendentry{$E_b/N_0 = 3.0$ dB}

\addplot[
    color=magenta,
    mark=pentagon,
    thick,
    mark size=3,
]
table {
1	0.0002619101
2	0.0001003933
3	0.000043427
4	1.96067415730337E-05
5	8.98876404494382E-06
6	5.28089887640449E-06
7	3.65168539325843E-06
8	3.25842696629213E-06
9	2.97752808988764E-06
10	2.86516853932584E-06
11	2.86516853932584E-06
12	2.86516853932584E-06
13	2.86516853932584E-06
14	0.000002809
15	0.000002809
16	0.000002809
17	0.000002809
18	0.000002809
19	0.000002809
20	0.000002809
};
\addlegendentry{$E_b/N_0 = 3.5$ dB - $\Omega=15$}

\addplot[
    color=magenta,
    dashed,
    mark=pentagon,
    thick,
    mark size=3,
]
table {
1	0.0002640223
2	9.36312849162011E-05
3	4.78212290502793E-05
4	2.8268156424581E-05
5	1.77653631284916E-05
6	1.18435754189944E-05
7	8.54748603351955E-06
8	0.000007095
9	5.53072625698324E-06
10	4.97206703910615E-06
11	4.46927374301676E-06
12	4.02234636871508E-06
13	3.79888268156425E-06
14	3.57541899441341E-06
15	3.29608938547486E-06
16	3.18435754189944E-06
17	3.18435754189944E-06
18	3.01675977653631E-06
19	2.84916201117318E-06
20	2.84916201117318E-06
};
\addlegendentry{$E_b/N_0 = 3.5$ dB}

\end{axis}
\end{tikzpicture}}
   \\
   \ref{TMAXvsFER_R05}
  \caption{FER curves for TSCF decoding (solid lines) and SC-Flip (dashed lines) with various $T_{max}$ values, for $PC(1024,512)$, $C=12$.}
  \label{fig:TMAXvsFER_R05}
\end{figure}
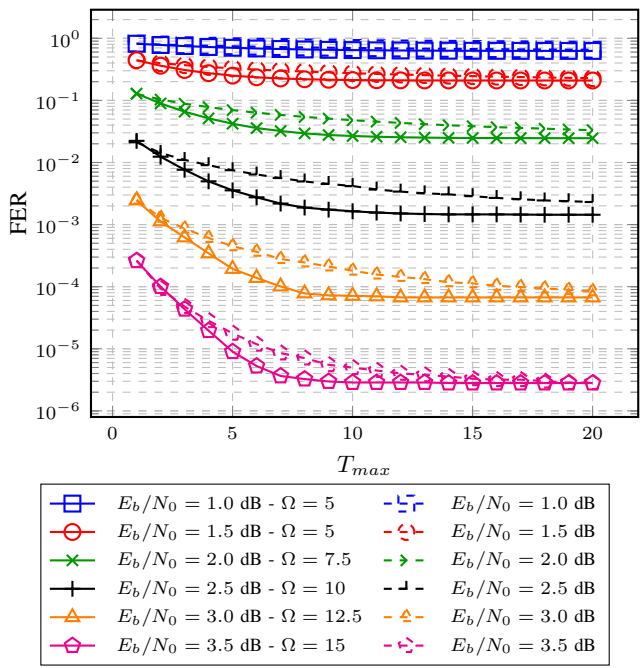

The advantages of TSCF decoding can also be observed in high and medium code rates. With increased rate, the size of $\boldsymbol{\mathcal{I}}^C$ increases, making it more difficult to identify the correct flipping index for both TSCF and SC-Flip decoding. Finally, $\boldsymbol{f^{E_1}}$ becomes more distributed over the whole codeword, as depicted in Fig.~\ref{fig:spikes2} for $PC(1024,512)$. Due to these reasons, the average number of iterations $T_{avg}$ increases with the rate, where a full iteration is considered as $N$ bits being decoded. Consequently, each decoding attempt by SC-Flip or TSCF accounts for a fraction of a full iteration. 

Table \ref{tab:numofindices} details the size of $\boldsymbol{\mathcal{I}^C}$ observed for different rates, for various $E_b/N_0$ points. As mentioned in Section~\ref{sec:indexselection}, the total number of critical indices for a fixed rate decreases with increasing $E_b/N_0$. On the other hand, as the rate increases, a higher number of non-frozen bits are present, and are assigned to less and less reliable channels. Consequently, more indices become susceptible to channel errors. In Table \ref{tab:numofindices_targetfer}, cardinality of critical sets with respect to a set of target FER ($\text{FER}_t \in \{ 10^{-2},10^{-3},10^{-4} \}$) for various rates are presented. Similar to Table \ref{tab:numofindices}, the size of the critical set increases with the code rate in general. On the other hand, as the target FER is improved, the cardinality of critical set decreases for any rate, since the occurrence of channel-induced errors decrease with increased $E_b/N_0$. 

\begin{table}[t!]
\centering
\caption{Number of critical indices with respect to $E_b/N_0$ and $R$ for $N=1024$. The optimal CRC length for each rate is chosen from Fig.~\ref{fig:CRC_Sweep_ALL}.}
\label{tab:numofindices}
\setlength{\extrarowheight}{1.25pt}
\begin{tabular}{clllll}
\toprule
$E_b/N_0$          & \multicolumn{5}{c}{Rate ($R$)} \\
\cmidrule(l){2-6}
{[}dB{]}           & 1/6  & 1/4  & 1/3  & 1/2 & 2/3 \\
\cmidrule(r){1-1}\cmidrule(l){2-6}
1.0                & 91  & 118  & 138 & 163 & 154 \\
1.5                & 89  & 110  & 131 & 160 & 169 \\
2.0                & 83  & 101  & 125 & 147 & 169 \\
2.5                & 72  & 85   & 106 & 130 & 155 \\
3.0                & 56  & 69   & 78  & 113 & 133 \\
3.5                & 39  & -    & -   & 74  & 109 \\
4.0                & -   & -    & -   & -   & 71  \\
\bottomrule
\end{tabular}
\end{table} 

\begin{table}[t!]
\centering
\caption{Number of critical indices with respect to target FER ($\text{FER}_t$) and $R$ for $N=1024$, $C=32$.}
\label{tab:numofindices_targetfer}
\setlength{\extrarowheight}{1.25pt}
\begin{tabular}{clllll}
\toprule
$\text{FER}_t$          & \multicolumn{5}{c}{Rate ($R$)} \\
\cmidrule(l){2-6}
          & 1/6  & 1/4  & 1/3  & 1/2 & 2/3 \\
\cmidrule(r){1-1}\cmidrule(l){2-6}
$10^{-2}$ & 64 & 86 & 114 & 131 & 126 \\
$10^{-3}$ & 47 & 71 & 98  & 117 & 107 \\
$10^{-4}$ & 32 & 58 & 76  & 93  & 80  \\
\bottomrule
\end{tabular}
\end{table} 

As the impact of $E_1$ decreases, the error-correction performance gap between baseline SC-Flip algorithm and SCO-1 narrows. As a result, the required $T_{max}$ value for SC-Flip algorithm to reach to the performance of SCO-1 decreases with increasing rate. In Fig.~\ref{fig:TMAXvsFER_R05} the FER performance of SC-Flip and TSCF are compared against each other for $PC(1024,512)$ and the optimal $\Omega$ value for each $E_b/N_0$. As $T_{max}$ increases, compared to $R = \frac{1}{6}$ (Fig.~\ref{fig:TMAXvsFER_R016}), the FER of both SC-Flip and TSCF converges faster to the lower limit imposed by the SCO-1 performance. Nonetheless, TSCF is shown to converge faster than SC-Flip at high rates, yielding comparatively improved error-correction performance and $T_{max}$.

Table~\ref{tab:Lvalues} presents the optimized $\Omega$ values for $N=1024$ and the rates considered in this work. $\Omega$ values are swept with a step size of $2.5$. It is observed that $\Omega$ values that provide the best error-correction performance increase with increasing $E_b/N_0$, since the LLR magnitude gap increases (Fig. \ref{fig:avLLRvsSNR}). On the other hand, it can be observed that the optimal $\Omega$ decreases as the code rate rises. This is due to the fact that the mean LLR magnitude in case of successful decoding decreases with the code rate.

\begin{table}[t!]
\centering
\caption{Optimal $\Omega$ values for $N=1024$ and various rates at different $E_b/N_0$ points.}
\label{tab:Lvalues}
\setlength{\extrarowheight}{1.25pt}
\begin{tabular}{clllll}
\toprule
$E_b/N_0$          & \multicolumn{5}{c}{Rate ($R$)} \\
\cmidrule(l){2-6}
{[}dB{]}           & 1/6  & 1/4  & 1/3  & 1/2 & 2/3 \\
\cmidrule(r){1-1}\cmidrule(l){2-6}
$1.0$ & $7.5$  & $5$    & $5$   & $5$    & $5$    \\
$1.5$ & $7.5$  & $7.5$  & $7.5$ & $5$    & $5$    \\
$2.0$ & $10$   & $10$   & $10$  & $7.5$  & $5$    \\
$2.5$ & $10$   & $10$   & $10$  & $10$   & $5$    \\
$3.0$ & $15$   & $12.5$ & $15$  & $12.5$ & $7.5$  \\
$3.5$ & $22.5$ & --	    & --    & $15$   & $10$   \\
$4.0$ & --     & --     & --    & --     & $12.5$ \\
\bottomrule
\end{tabular}
\end{table}

\section{Simulation Results}\label{sec:simresults}

In this Section, we evaluate the error-correction performance, the maximum number of iterations $T_{max}$, and average number of iterations $T_{avg}$ of TSCF decoding, with respect to SC-Flip, SC-Oracle and SC-List decoding. For all SC-Flip based implementations including TSCF, $C \in \{7,9,9,12,12\}$ for $R \in \{\frac{1}{6},\frac{1}{4},\frac{1}{3},\frac{1}{2},\frac{2}{3}\}$ are selected as CRC remainder length, using Fig.~\ref{fig:CRC_Sweep_ALL}. While an outer CRC code improves the error-correction performance of SC-List decoding significantly at medium to high code rates, it was shown in \cite{eMBBcode-Asilomar17} that at low code rates the CRC is detrimental. Thus, for a fair comparison, we did not consider any CRC in the SC-List curves for low-rate codes. For the performance comparison of high-rate codes, CRC-aided SC-List is used with $C=8$ \cite{eMBBcode-Asilomar17}.

\subsection{Error-Correction Performance}

\begin{figure}[t!]
  \centering
   \scalebox{1}{\begin{tikzpicture}[spy using outlines=
	{rectangle, magnification=2, connect spies}]
  \pgfplotsset{
    label style = {font=\fontsize{9pt}{7.2}\selectfont},
    tick label style = {font=\fontsize{7pt}{7.2}\selectfont}
  }

\begin{axis}[
	scale = 1,
    ymode=log,
    xlabel={$E_b/N_0$ [\text{dB}]}, xlabel style={yshift=0.4em},
    ylabel={FER}, ylabel style={yshift=-0.75em},
    grid=both,
    ymajorgrids=true,
    xmajorgrids=true,
    grid style=dashed,
    width=\columnwidth, height=7cm,
    thick,
    mark size=3,
    legend cell align={left},
    legend to name=FER-LMULT,
    legend columns=2,
]

\addplot[
    color=black,
    dashed,
    mark=x,
    thick,
    mark size=3,
]
table {
1.0	6.27250e-1
1.5	3.75080e-1
2.0	1.68380e-1
2.5	5.32700e-2
3.0	1.10000e-2
3.5	1.52000e-3
};
\addlegendentry{SC}

\addplot[
    color=red,
    mark=triangle,
    thick,
    mark size=3,
]
table {
1.0	4.81090e-1
1.5	2.33640e-1
2.0	7.90500e-2
2.5	1.86100e-2
3.0	3.10000e-3
3.5	4.38504e-4
};
\addlegendentry{SC-Flip \cite{SCFlip14}}

\addplot[
    color=black,
    mark=diamond,
    thick,
    mark size=3,
]
table {
1.0	4.06690e-1
1.5	1.70010e-1
2.0	4.37500e-2
2.5	7.35000e-3
3.0	7.10000e-4
3.5	5.11345e-5
};
\addlegendentry{SC-Flip-EIS \cite{SCF-WCNC18}}

\addplot[
    color=blue,
    mark=o,
    thick,
    mark size=3,
]
table {
1.0	3.82800e-1
1.5	1.54740e-1
2.0	3.92400e-2
2.5	6.21000e-3
3.0	5.50000e-4
3.5	3.47142e-5
};
\addlegendentry{TSCF}

\addplot[
    color=red,
    mark=+,
    thick,
    mark size=3,
]
table {
1.0 3.53100e-01
1.5 1.45200e-01
2.0 3.93000e-02
2.5 7.50000e-03
3.0 5.99003e-04
3.5 4.08481e-05
};
\addlegendentry{SC-List ($L=2$)}

\addplot[
    color=green!60!black,
    mark=pentagon,
    thick,
    mark size=3,
]
table {
1.0 1.98300e-01
1.5 6.29000e-02
2.0 1.00000e-02
2.5 1.14574e-03
3.0 5.84527e-05
3.5 1.93510e-06 
};
\addlegendentry{SC-List $L=4$}

\addplot[
    color=magenta,
    mark=square,
    thick,
    mark size=3,
]
table {
1.0	3.52300e-1
1.5	1.34300e-1
2.0	3.24300e-2
2.5	4.49000e-3
3.0	3.30000e-4
3.5	2.15783e-5
};
\addlegendentry{SCO-1}

\coordinate (spypoint) at (axis cs:3.0,2e-4);
\coordinate (magnifyglass) at (axis cs:1.45,2e-4);

\end{axis}
\spy [magenta, height=2.6cm, width=2.2cm] on (spypoint)
   in node[fill=white] at (magnifyglass);
\end{tikzpicture}}
   \\
   \ref{FER-LMULT}
  \caption{FER curves with various decoding algorithms for $PC(1024,170)$. $C=7$ and $T_{max} = 10$ for all SC-Flip based implementations.}
  \label{fig:FER-LMULT}
\end{figure}

The error-correction performance of TSCF is compared to SC, SC-Flip, SC-Flip-EIS from \cite{SCF-WCNC18}, SC-List and SCO-1. Simulations are performed with binary phase-shift keying (BPSK) modulation and additive white Gaussian noise (AWGN) channel. The same $T_{max}$ is selected for all SC-Flip based algorithms, the lowest value for which TSCF decoding meets the error-correction performance of SCO-1, as a $T_{max}$ higher than that will not improve the FER any further. Fig.~\ref{fig:FER-LMULT} presents the error-correction performance for $PC(1024,170)$. 

Multiple $\Omega$ values are used from Table~\ref{tab:Lvalues} in order to optimize the error-correction performance for each $E_b/N_0$ point. Set $\boldsymbol{\mathcal{I}^C}$ for $E_b/N_0 = 3.0$ dB from Table~\ref{tab:numofindices} is used. It can be seen that TSCF decoding achieves the SCO-1 performance with $T_{max} = 10$, outperforming SC-Flip decoding with the same $T_{max}$ value by $0.43$ dB at FER of $10^{-4}$. SC-Flip-EIS decoding from \cite{SCF-WCNC18} has an error-correction performance that also approaches to SCO-1 performance, but it can be seen that it is slightly worse than that of TSCF. It can also be seen that TSCF decoding approaches SC-List decoding performance with list size $L=2$.

\begin{figure}[t!]
  \centering
   \scalebox{1}{\begin{tikzpicture}[spy using outlines=
	{rectangle, magnification=2, connect spies}]
  \pgfplotsset{
    label style = {font=\fontsize{9pt}{7.2}\selectfont},
    tick label style = {font=\fontsize{7pt}{7.2}\selectfont}
  }

\begin{axis}[
	scale = 1,
    ymode=log,
    xlabel={$E_b/N_0$ [\text{dB}]}, xlabel style={yshift=0.4em},
    ylabel={FER}, ylabel style={yshift=-0.75em},
    grid=both,
    ymajorgrids=true,
    xmajorgrids=true,
    grid style=dashed,
    width=\columnwidth, height=7cm,
    thick,
    mark size=3,
    legend cell align={left},
    legend to name=FER-LSING,
    legend columns=2,
]

\addplot[
    color=black,
    dashed,
    mark=x,
    thick,
    mark size=3,
]
table {
1.0	6.27250e-1
1.5	3.75080e-1
2.0	1.68380e-1
2.5	5.32700e-2
3.0	1.10000e-2
3.5	1.52000e-3
};
\addlegendentry{SC}

\addplot[
    color=red,
    mark=triangle,
    thick,
    mark size=3,
]
table {
1.0	4.81090e-1
1.5	2.33640e-1
2.0	7.90500e-2
2.5	1.86100e-2
3.0	3.10000e-3
3.5	4.38504e-4
};
\addlegendentry{SC-Flip \cite{SCFlip14}}

\addplot[
    color=blue,
    mark=o,
    thick,
    mark size=3,
]
table {
1.0	3.82630e-1
1.5	1.54130e-1
2.0	4.26800e-2
2.5	7.70000e-3
3.0	1.20000e-3
3.5	1.14013e-4
};
\addlegendentry{TSCF $\Omega=7.5$}

\addplot[
    color=black,
    mark=+,
    thick,
    mark size=3,
]
table {
1.0	4.83020e-1
1.5	2.09520e-1
2.0	5.04800e-2
2.5	6.69000e-3
3.0	4.98300e-4
3.5	4.43356e-5
};
\addlegendentry{TSCF $\Omega=15$}

\addplot[
    color=green!60!black,
    mark=pentagon,
    thick,
    mark size=3,
]
table {
1.0	5.26350e-1
1.5	2.68190e-1
2.0	8.73100e-2
2.5	1.43700e-2
3.0	1.03000e-3
3.5	5.04130e-5
};
\addlegendentry{TSCF $\Omega=22.5$}

\addplot[
    color=magenta,
    mark=square,
    thick,
    mark size=3,
]
table {
1.0	3.52300e-1
1.5	1.34300e-1
2.0	3.24300e-2
2.5	4.49000e-3
3.0	3.30000e-4
3.5	2.15783e-5
};
\addlegendentry{SCO-1}

\coordinate (spypoint) at (axis cs:3.1,2e-4);
\coordinate (magnifyglass) at (axis cs:1.45,2e-4);

\end{axis}
\spy [magenta, height=2.6cm, width=2.2cm] on (spypoint)
   in node[fill=white] at (magnifyglass);
\end{tikzpicture}}
   \\
   \ref{FER-LSING}
  \caption{FER curves with various decoding algorithms for $PC(1024,170)$. $C=7$ and $T_{max} = 10$ for all SC-Flip based implementations.}
  \label{fig:FER-LSING}
\end{figure}

If a single $\Omega$ is allowed regardless of the $E_b/N_0$ value, it should be selected from Table~\ref{tab:Lvalues} to optimize the error-correction performance at ${\rm FER}_t$. Fig.~\ref{fig:FER-LSING} presents the FER of TSCF decoding with single $\Omega$ values. It can be seen that while some $\Omega$ values help TSCF match with SCO-1 performance at low $E_b/N_0$, the FER diverges towards SC-Flip performance as $E_b/N_0$ increases. On the other hand, if an $\Omega$ value for a higher $E_b/N_0$ is selected, its performance diverges from SCO-1 as the $E_b/N_0$ decreases. Finally, a moderate $\Omega$ value can maintain an acceptable FER for both low and high $E_b/N_0$ points; in case of Fig.~\ref{fig:FER-LSING}, $\Omega = 15$ yields an error-correction performance that is close to the SCO-1 performance at all considered $E_b/N_0$ values.

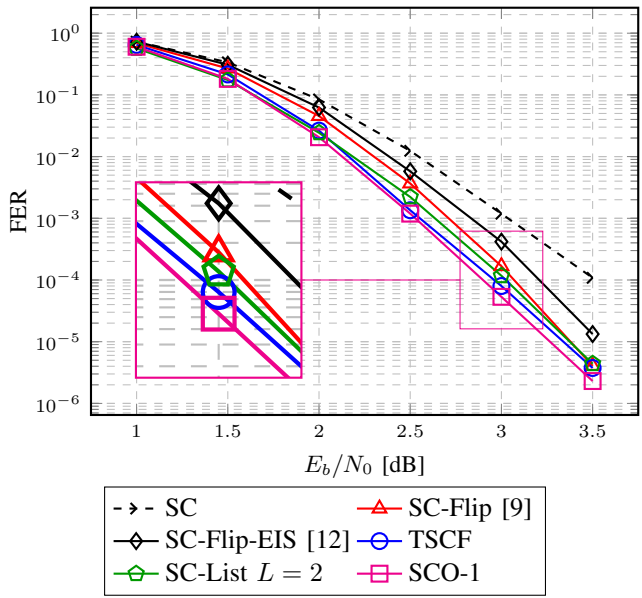
\begin{figure}[t!]
  \centering
   \scalebox{1}{\begin{tikzpicture}[spy using outlines=
	{rectangle, magnification=2, connect spies}]
  \pgfplotsset{
    label style = {font=\fontsize{9pt}{7.2}\selectfont},
    tick label style = {font=\fontsize{7pt}{7.2}\selectfont}
  }

\begin{axis}[
	scale = 1,
    ymode=log,
    xlabel={$E_b/N_0$ [\text{dB}]}, xlabel style={yshift=0.4em},
    ylabel={FER}, ylabel style={yshift=-0.75em},
    grid=both,
    ymajorgrids=true,
    xmajorgrids=true,
    grid style=dashed,
    width=\columnwidth, height=7cm,
    thick,
    mark size=3,
    legend cell align={left},
    legend to name=FER-LMULT-k512,
    legend columns=2,
]

\addplot[
    color=black,
    dashed,
    mark=x,
    thick,
    mark size=3,
]
table {
1.0 7.28900e-01
1.5 3.36900e-01
2.0 8.44000e-02
2.5 1.24000e-02
3.0 1.18618e-03
3.5 1.08427e-04
};
\addlegendentry{SC}

\addplot[
    color=red,
    mark=triangle,
    thick,
    mark size=3,
]
table {
1.0	6.90840e-1
1.5	2.67550e-1
2.0	4.54700e-2
2.5	3.67000e-3
3.0	1.67497e-4
3.5	4.05281e-6
};
\addlegendentry{SC-Flip \cite{SCFlip14}}

\addplot[
    color=black,
    mark=diamond,
    thick,
    mark size=3,
]
table {
1.0	7.27860e-1
1.5	3.08090e-1
2.0	6.23600e-2
2.5	5.77000e-3
3.0	4.17924e-4
3.5	1.32886e-5
};
\addlegendentry{SC-Flip-EIS \cite{SCF-WCNC18}}

\addplot[
    color=blue,
    mark=o,
    thick,
    mark size=3,
]
table {
1.0	6.45120e-1
1.5	2.14410e-1
2.0	2.63400e-2
2.5	1.35000e-3
3.0	7.98968e-5
3.5	3.74981e-6
};
\addlegendentry{TSCF}

\addplot[
    color=green!60!black,
    mark=pentagon,
    thick,
    mark size=3,
]
table {
1.0 0.5668
1.5 0.1742
2.0 0.0243
2.5 0.00221572
3.0 0.000117092
3.5 4.39747e-06
};
\addlegendentry{SC-List $L=2$}

\addplot[
    color=magenta,
    mark=square,
    thick,
    mark size=3,
]
table {
1.0	6.02980e-1
1.5	1.81870e-1
2.0	2.06100e-2
2.5	1.19000e-3
3.0	5.33590e-5
3.5	2.30344e-6
};
\addlegendentry{SCO-1}

\coordinate (spypoint) at (axis cs:3.0,1e-4);
\coordinate (magnifyglass) at (axis cs:1.45,1e-4);

\end{axis}
\spy [magenta, height=2.6cm, width=2.2cm] on (spypoint)
   in node[fill=white] at (magnifyglass);
\end{tikzpicture}}
   \\
   \ref{FER-LMULT-k512}
  \caption{FER curves with various decoding algorithms for $PC(1024,512)$. $C = 12$ for all SC-Flip based decoders including SCO-1, $C = 8$ for SC-List decoder, and $T_{max} = 10$ for SC-Flip, SC-Flip-EIS and TSCF decoding.}
  \label{fig:FER-LMULT-k512}
\end{figure}

Fig.~\ref{fig:FER-LMULT-k512} depicts the error-correction performance of TSCF decoding against SC, SC-Flip, SC-Flip-EIS \cite{SCF-WCNC18} and SC-List ($L=2$) decoding for $PC(1024,512)$. The CRC remainder length is chosen to be $C=8$ for SC-List decoder using \cite{eMBBcode-Asilomar17}, while $C=12$ and $T_{max}=10$ for all SC-Flip based decoders. Compared to Fig.~\ref{fig:FER-LMULT} where the performance comparison for low-rate codes is demonstrated, the performance gap between SC-Flip and SCO-1 FER curves is smaller. This is due to the fact that the number of $E_1$ errors decreases as the rate increases, as highlighted earlier in Fig.~\ref{fig:errorlikelihoods-rate}. It can be seen that TSCF has superior error-correction performance compared to SC-Flip and SC-Flip-EIS, as it is closest to SCO-1 performance. Moreover, the curve obtained SC-List with $L=2$ matches that of TSCF. Finally, SC-Flip-EIS decoding has a worse error-correction performance than SC-Flip: the effectiveness of the criteria described in \cite{SCF-WCNC18} degrades as the code rate increases, since the size of critical set increases with rate and SC-Flip-EIS does not use any further restrictions to reduce the search space for the flipping index.

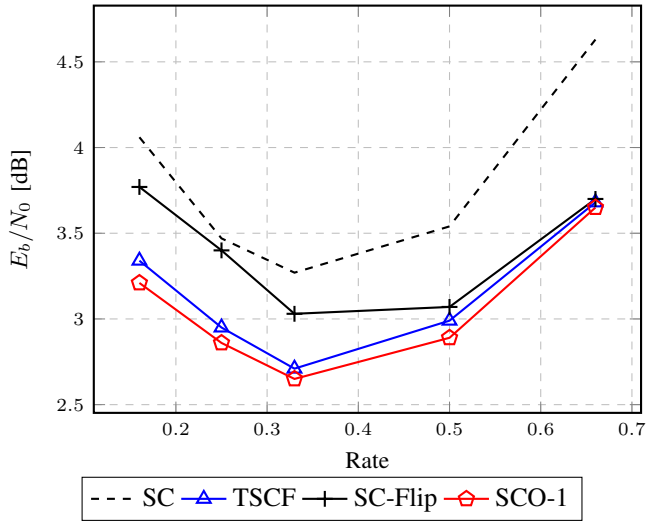
\begin{figure}
  \centering
   \scalebox{1}{\begin{tikzpicture}[spy using outlines=
	{rectangle, magnification=2, connect spies}]
  \pgfplotsset{
    label style = {font=\fontsize{9pt}{7.2}\selectfont},
    tick label style = {font=\fontsize{7pt}{7.2}\selectfont}
  }

\begin{axis}[
	scale = 1,
    xlabel={Rate}, xlabel style={yshift=0.4em},
    ylabel={$E_b/N_0$ [\text{dB}]}, ylabel style={yshift=-0.75em},
    grid=both,
    ymajorgrids=true,
    xmajorgrids=true,
    grid style=dashed,
    width=\columnwidth, height=7cm,
    thick,
    mark size=3,
    legend to name=EbN0vsR,
    legend columns=4,
]

\addplot[
    color=black,
    dashed,
    thick,
]
table {
0.16	4.06
0.25	3.47
0.33	3.27
0.50	3.54
0.66	4.63
};
\addlegendentry{SC}

\addplot[
    color=blue,
    mark=triangle,
    thick,
    mark size=3,
]
table {
0.16	3.34
0.25	2.95
0.33	2.71
0.50	2.99
0.66	3.68
};
\addlegendentry{TSCF}

\addplot[
    color=black,
    mark=+,
    thick,
    mark size=3,
]
table {
0.16	3.77
0.25	3.40
0.33	3.03
0.50	3.07
0.66	3.70
};
\addlegendentry{SC-Flip}

\addplot[
    color=red,
    mark=pentagon,
    thick,
    mark size=3,
]
table {
0.16	3.21
0.25	2.86
0.33	2.65
0.50	2.89
0.66	3.65
};
\addlegendentry{SCO-1}

\end{axis}
\end{tikzpicture}}
   \\
   \ref{EbN0vsR}
  \caption{$E_b/N_0$ requirements for different decoding approaches to achieve ${\rm FER}_t=10^{-4}$, for $N=1024$, $\Omega \in \{15,15,12.5,10,10\}$ and $T_{max} = 10$ for $R \in \{\frac{1}{6}, \frac{1}{4}, \frac{1}{3}, \frac{1}{2}, \frac{2}{3}\}$, respectively.}
  \label{fig:EbN0vsR}
\end{figure}

Fig.~\ref{fig:EbN0vsR} presents the $E_b/N_0$ requirements to achieve target ${\rm FER}_t=10^{-4}$ for $N=1024$, $R \in \{\frac{1}{6}, \frac{1}{4}, \frac{1}{3}, \frac{1}{2}, \frac{2}{3}\}$ under different decoding approaches. For TSCF decoding, a single $\Omega$ is selected for each rate from Table~\ref{tab:Lvalues}, regardless of the $E_b/N_0$ value. According to Fig.~\ref{fig:EbN0vsR}, TSCF decoding performs very close to the SCO-1 in all cases. As mentioned in Section~\ref{sec:highrate}, the FER of SC-Flip approaches to that of SCO-1 as the rate increases. On the other hand, the benefit of TSCF decoding in terms of error-correction performance is maximum at $R = \frac{1}{4}$, with an improvement of $0.45$ dB compared to baseline SC-Flip decoding. Finally, compared to SC decoding, TSCF decoding has the highest gain at $R = \frac{2}{3}$ ($0.95$ dB).

\subsection{Complexity}

The (average) computational complexity of SC-Flip is directly proportional to the (average) number of iterations. A single iteration of SC-Flip consists of SC and CRC decoding, and an additional selection and sorting process for the flipping indices based on their LLR value when the initial CRC decoding fails. On the other hand, an iteration of TSCF is composed of SC+CRC decoding, and a comparison for critical indices to identify whether the associated LLR is lower than $\Omega$. In this Section, we compare $T_{avg}$ for both algorithms at matching $T_{max}$ and at matching FER. 

Fig.~\ref{fig:Tave_matchedPerf} presents the average number of iterations for both SC-Flip and TSCF decoding, for $R = \frac{1}{6}$ and $R = \frac{1}{2}$ at matching FER=$10^{-4}$. $T_{max}$ of SC-Flip is kept at $10$ and $T_{max}$ of TSCF is tuned to match the error-correction performance of SC-Flip. For $R = \frac{1}{6}$ ($R = \frac{1}{2}$), the FER of TSCF matches that of SC-Flip when $T_{max} = 3$ ($T_{max} = 5$). As $E_b/N_0$ is increased, the average number of iterations converges to $1$, regardless of rate and decoding algorithm. In all cases, TSCF converges to $T_{avg}=1$ quicker than SC-Flip algorithm, with up to $14\%$ lower $T_{avg}$ for $R = \frac{1}{6}$ and $R = \frac{1}{2}$. 

\begin{figure}
  \centering
   \scalebox{1}{\begin{tikzpicture}
  \pgfplotsset{
    label style = {font=\fontsize{9pt}{7.2}\selectfont},
    tick label style = {font=\fontsize{7pt}{7.2}\selectfont}
  }

\begin{axis}[
	scale = 1,
    xlabel={$E_b/N_0$ [\text{dB}]}, xlabel style={yshift=0.4em},
    ylabel={Time Steps}, ylabel style={yshift=-0.15em},
    grid=both,
    ymajorgrids=true,
    xmajorgrids=true,
    grid style=dashed,
    width=\columnwidth, height=7cm,
    thick,
    mark size=3,
    legend style={
      anchor={center},
      cells={anchor=west},
      column sep= 2mm,
      font=\fontsize{7pt}{7.2}\selectfont,
    },
    legend to name=timeComplexity,
    legend columns=2,
]

\addplot[
    color=red,
    mark=square,
    thick,
    mark size=3,
]
table {
1	8388.6
1.5	4746.72
2	2618.88
2.5	2107.38
3	2046
3.5	2046
};
\addlegendentry{SC-Flip ($T_{max}=10$)}

\addplot[
    color=blue,
    mark=triangle,
    thick,
    mark size=3,
]
table {
1	7263.3
1.25	5994.78
1.5	4664.88
1.75	3457.74
2	2721.18
2.25	2291.52
2.5	2127.84
2.75	2066.46
3	2046
3.25	2046
3.5	2046
};
\addlegendentry{TSCF ($T_{max}=5$)}

\addplot[
    color=magenta,
    mark=pentagon,
    thick,
    mark size=3,
]
table {
1.00	3939
1.25	3350
1.50	2941
1.75	2711
2.00	2609
2.25	2558
2.50	2558
2.75	2558
3.00	2558
3.25	2558
3.50	2558
};
\addlegendentry{Adaptive SC-List ($L_{max}=2$)}

\addplot[
    color=black,
    mark=o,
    thick,
    mark size=3,
]
table {
1.0 2046
1.5 2046
2.0 2046
2.5 2046
3.0 2046
3.5 2046
};
\addlegendentry{SC}

\addplot[
    color=green!60!black,
    mark=x,
    thick,
    mark size=3,
]
table {
1.0 2558
1.5 2558
2.0 2558
2.5 2558
3.0 2558
3.5 2558
};
\addlegendentry{SC-List}

\end{axis}
\end{tikzpicture}}
   \\
   \ref{timeComplexity}
  \caption{Time complexity comparison of SC-Flip and TSCF for $PC(1024,512)$ with SC, SC-List and Adaptive SC-List decoders.} 
  \label{fig:timeComplexity}
\end{figure}
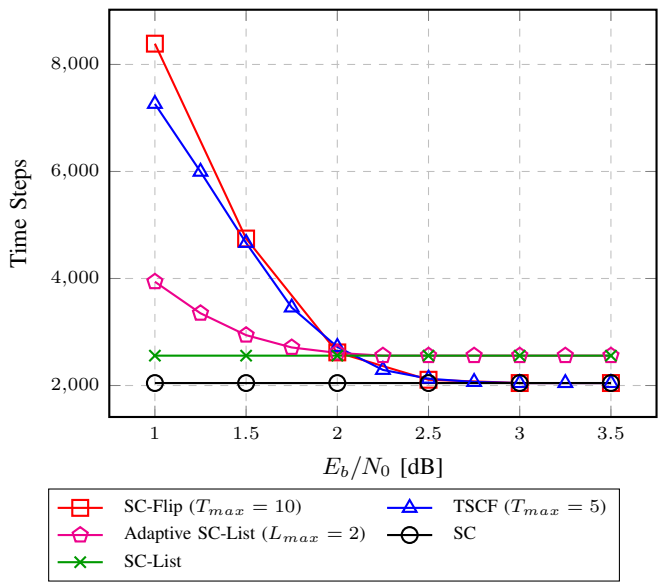

Fig.~\ref{fig:timeComplexity} compares the time complexity of SC-Flip and TSCF decoders with SC, SC-List and Adaptive SC-List decoders, for $PC(1024,512)$. According to \cite{arikan09}, $2N-2$ time steps are needed to complete a full SC tree traversal. Consequently, both TSCF and SC-Flip implementations converge to $2046$ average time steps with increasing $E_b/N_0$. On the other hand, $2N+K-2$ time steps are needed to complete one iteration of SC-List decoding \cite{list-LLR}. Thus, although SC-Flip and TSCF take more time steps on average at low $E_b/N_0$ values, they consume less time after $E_b/N_0 = 2$ dB. Compared to Adaptive SC-List \cite{adaptiveSCL2012}, TSCF has up to $83\%$ more time steps; however the gap is closed quickly with growing $E_b/N_0$. At moderate-to-high $E_b/N_0$, TSCF is shown to have $25\%$ less time steps than Adaptive SC-List. Finally, at matched FER, TSCF requires up to $14\%$ less time steps than SC-Flip implementation on average.

\begin{figure}
  \centering
   \scalebox{1}{\begin{tikzpicture}
  \pgfplotsset{
    label style = {font=\fontsize{9pt}{7.2}\selectfont},
    tick label style = {font=\fontsize{7pt}{7.2}\selectfont}
  }

\begin{axis}[
	scale = 1,
    xlabel={$E_b/N_0$ [\text{dB}]}, xlabel style={yshift=0.4em},
    ylabel={$T_{avg}$}, ylabel style={yshift=-0.75em},
    grid=both,
    ymajorgrids=true,
    xmajorgrids=true,
    grid style=dashed,
    width=\columnwidth, height=7cm,
    thick,
    mark size=3,
    legend style={
      anchor={center},
      cells={anchor=west},
      column sep= 2mm,
      font=\fontsize{7pt}{7.2}\selectfont,
    },
    legend to name=Tave_matchedPerf,
    legend columns=2,
]

\addplot[
    color=blue,
    mark=o,
    thick,
    mark size=3,
]
table {
1.0	2.21
1.5	1.66
2.0	1.27
2.5	1.01
3.0	1.00
3.5	1.00
};
\addlegendentry{SC-Flip $R = \frac{1}{6},T_{max}=10$}

\addplot[
    color=red,
    mark=x,
    thick,
    mark size=3,
]
table {
1.00	1.89
1.25	1.73
1.50	1.55
1.75	1.39
2.00	1.24
2.25	1.14
2.50	1.07
2.75	1.04
3.00	1.01
3.25	1.00
3.50	1.00
};
\addlegendentry{TSCF $R = \frac{1}{6},T_{max}=3$}

\addplot[
    color=green!60!black,
    mark=square,
    thick,
    mark size=3,
]
table {
1.0	4.10
1.5	2.32
2.0	1.28
2.5	1.03
3.0	1.00
3.5	1.00
};
\addlegendentry{SC-Flip $R = \frac{1}{2},T_{max}=10$}

\addplot[
    color=magenta,
    mark=triangle,
    thick,
    mark size=3,
]
table {
1.00	3.55
1.25	2.93
1.50	2.28
1.75	1.69
2.00	1.33
2.25	1.12
2.50	1.04
2.75	1.01
3.00	1.00
3.25	1.00
3.50	1.00
};
\addlegendentry{TSCF $R = \frac{1}{2},T_{max}=5$}

\addplot[
    color=black,
    thick,
    mark size=3,
]
table {
1.0 1
1.5 1
2.0 1
2.5 1
3.0 1
3.5 1
};

\addplot[
    color=black,
    dashed,
    thick,
    mark size=3,
]
table {
1.0 2
1.5 2
2.0 2
2.5 2
3.0 2
3.5 2
};

\addplot[
    color=black,
    dotted,
    thick,
    mark size=3,
]
table {
1.0 4
1.5 4
2.0 4
2.5 4
3.0 4
3.5 4
};

\node[] at (axis cs: 3.05,2.1) {SC-List $L = 2$};
\node[] at (axis cs: 3.05,4.1) {SC-List $L = 4$};

\end{axis}
\end{tikzpicture}}
   \\
   \ref{Tave_matchedPerf}
  \caption{$T_{avg}$ for SC-Flip and TSCF, $R = \{\frac{1}{6} ,\frac{1}{2}\}$, at matched FER.}
  \label{fig:Tave_matchedPerf}
\end{figure}
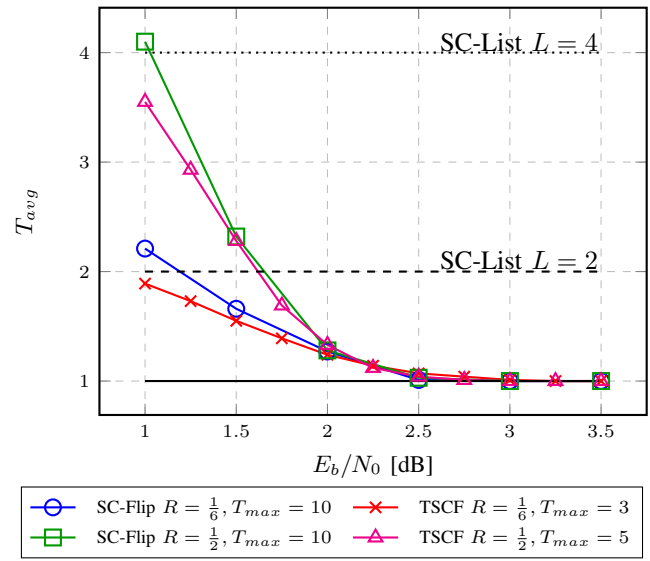

\begin{table}[t!]
\centering
\caption{$T_{avg}^{max}$ values for TSCF and SC-Flip at matching $T_{max} = 10$.}
\label{tab:matchedTmax}
\setlength{\extrarowheight}{1.25pt}
\begin{tabular}{lccccc}
\toprule
 Rate ($R$)& $1/6$ & $1/4$ & $1/3$ & $1/2$ & $2/3$\\
\cmidrule(r){1-1} \cmidrule(l){2-6}
 SC-Flip  & $2.21$ & $2.42$ & $2.59$ & $4.10$ & $6.55$ \\
 TSCF     & $3.02$ & $2.95$ & $3.04$ & $5.18$ & $8.20$ \\
\bottomrule
\end{tabular}
\end{table} 

\begin{table}[t!]
\centering
\caption{Observed $T_{max}$ and worst case average number of iterations ($T_{avg}^{max}$) values for TSCF at matched error-correction performance with SC-Flip decoding at FER = $10^{-4}$.}
\label{tab:Tvalues-SCF}
\setlength{\extrarowheight}{1.25pt}
\begin{tabular}{llccccc}
\toprule

& Rate & $1/6$ & $1/4$ & $1/3$ & $1/2$ & $2/3$\\
\cmidrule(r){1-2} \cmidrule(l){3-7}
\multirow{2}{*}{SC-Flip} & 
 $T_{max}$     & $10$   & $10$   & $10$   & $10$   & $10$ \\
&$T_{avg}^{max}$  & $2.21$ & $2.42$ & $2.59$ & $4.10$ & $6.55$ \\
\cmidrule{1-7}
\multirow{3}{*}{TSCF} & 
 $\Omega$ & $15$   & $12.5$   & $15$ & $15$   & $12.5$ \\
&$T_{max}$     & $3$    & $2$    & $3$    & $5$    & $7$ \\
&$T_{avg}^{max}$  & $1.80$ & $1.45$ & $1.97$ & $3.53$ & $6.13$ \\
\bottomrule
\end{tabular}
\end{table} 

Table~\ref{tab:Tvalues-SCF} reports $T_{max}$ and the maximum $T_{avg}$ observed among the considered $E_b/N_0$ points, denoted by $T_{avg}^{max}$. The $\Omega$ values are selected from Table~\ref{tab:Lvalues}, according to Fig.~\ref{fig:EbN0vsR}. The FER of both algorithms matches in all cases. According to the results, at FER=$10^{-4}$ TSCF has both lower $T_{max}$ (up to $5\times$) and $T_{avg}$ (up to $40\%$) at the same time. For matching $T_{max}$, while TSCF outperforms SC-Flip in terms of FER, it has higher $T_{avg}$, as reported in Table~\ref{tab:matchedTmax}. For example, $T_{avg}$ for TSCF is $26\%$ higher than that of SC-Flip at $E_b/N_0 = 1$ dB when $T_{max} = 10$ for $R = \frac{1}{2}$. This is due to the fact that TSCF tends to flip the indices located towards the left side of the polar code, where SC-Flip does not have such a constraint. Thus, each decoding attempt by TSCF will averagely re-decode more bits and contribute more to the total $T_{avg}$.

\begin{table}[t!]
\centering
\caption{$T_{max}$ values and $E_b/N_0$ gain at FER~=~$10^{-4}$ for TSCF compared to SC-Flip at matching $T_{avg}^{max}$ and SC-Flip $T_{max}=10$.}
\label{tab:matchedTave}
\setlength{\extrarowheight}{1.25pt}
\begin{tabular}{lccccc}
\toprule
 Rate ($R$) & $1/6$ & $1/4$ & $1/3$ & $1/2$ & $2/3$\\
 \cmidrule(r){1-1} \cmidrule(l){2-6}
 $T_{max}$  & $4$     & $4$     & $5$     & $6$     & $7$ \\
 $E_b/N_0$ Gain (dB)  & $0.37$ & $0.39$ & $0.28$ & $0.05$ & $0.01$ \\
\bottomrule
\end{tabular}
\end{table} 

Table~\ref{tab:matchedTave} presents the $T_{max}$ values and $E_b/N_0$ gain for TSCF for when $T_{avg}^{max}$ of SC-Flip and TSCF match at all $E_b/N_0$ values. $T_{max}$ for SC-Flip is kept constant at $10$. Under these constraints, depending on the code rate, TSCF has up to $2.5\times$ lower worst case latency than SC-Flip, with a concurrent $E_b/N_0$ gain up to $0.39$ dB.

TSCF shows improved error-correction performance and reduced $T_{max}$ requirements when compared to the SCF-EIS and SCF-FIS criteria in \cite{SCF-WCNC18}. Compared to the technique in \cite{SCF-GLOBECOM17} targeting SCO-1, our methodology approaches the SCO-1 performance with significantly lower number of $T_{max}$. Similarly, when compared to \cite{SCFlip17-jour} targeting SCO-1, TSCF decoding yields similar error-correction performance at lower $T_{max}$. Finally, TSCF has comparable error-correction performance to the PSCF algorithm from \cite{PSCF-ICC18}, with lower $T_{max}$ requirements.

\section{Conclusion} \label{sec:conc}

In this work, we have presented a thresholded index selection criterion that can be applied to SC-Flip decoding of polar codes, creating the thresholded SC-Flip (TSCF) algorithm. It is based on the identification of a set of error-prone bit indices and of an LLR threshold to determine the position of a channel-induced wrong bit estimation. We show how to optimize the threshold value, and how to efficiently select the erroneous indices. It is estimated to reduce the implementation complexity of SC-Flip. Simulations are performed over a wide range of rates, and demonstrate negligible performance loss for TSCF decoding compared to SCO-1 performance, with less than $10$ maximum number of iterations. At matching maximum number of iterations, TSCF decoding has up to $0.45$ dB gain in FER, compared to SC-Flip. At matching FER=$10^{-4}$, TSCF requires up to $5\times$ lower maximum number of iterations and has up to $40\%$ lower average iterations. At matching $T_{avg}^{max}$, TSCF requires up to $60\%$ lower maximum number of iterations, with an $E_b/N_0$ gain of up to $0.39$ dB. Given the reduced computational complexity, reduced number of iterations and improved error-correction performance, TSCF is a good polar code decoding algorithm candidate for upcoming 5G applications.

\section*{Acknowledgements}
The authors would like to thank Seyyed Ali Hashemi and Harsh Aurora for their valuable contributions to this work.


\begin{IEEEbiography}[{\includegraphics[width=1in,height=1.25in,clip,keepaspectratio]{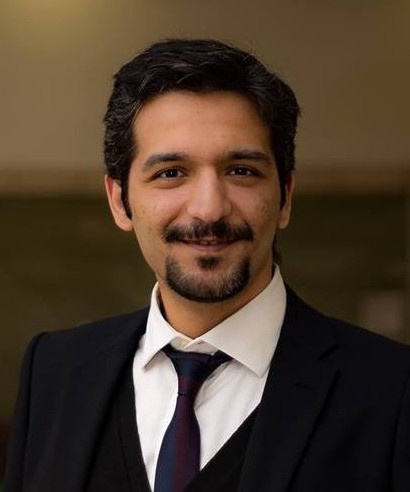}}]{Furkan Ercan}
(S'11) received the B.Sc. degree in Electrical and Electronics Engineering in 2011 and M.Sc. degree in Sustainable Environment and Energy Systems in 2015, both from Middle East Technical University (METU) Northern Cyprus Campus (NCC), Ankara, Turkey. From 2011 to 2012, he worked as a full time R\&D intern at Intel Corporation in Hillsboro, OR, USA, focusing on system-level energy efficiency on enterprise platforms. He is currently pursuing a Ph.D. degree with McGill University, Montr\'eal, QC, Canada. His research interests are algorithm, design and implementation of signal processing systems with a focus on polar codes, and energy aware hardware architectures. He received a Best Student Paper Award in 2015 IEEE International Conference in Energy Aware Computing (ICEAC) in Cairo, Egypt.
\end{IEEEbiography}

\begin{IEEEbiography}[{\includegraphics[width=1in,height=1.25in,clip,keepaspectratio]{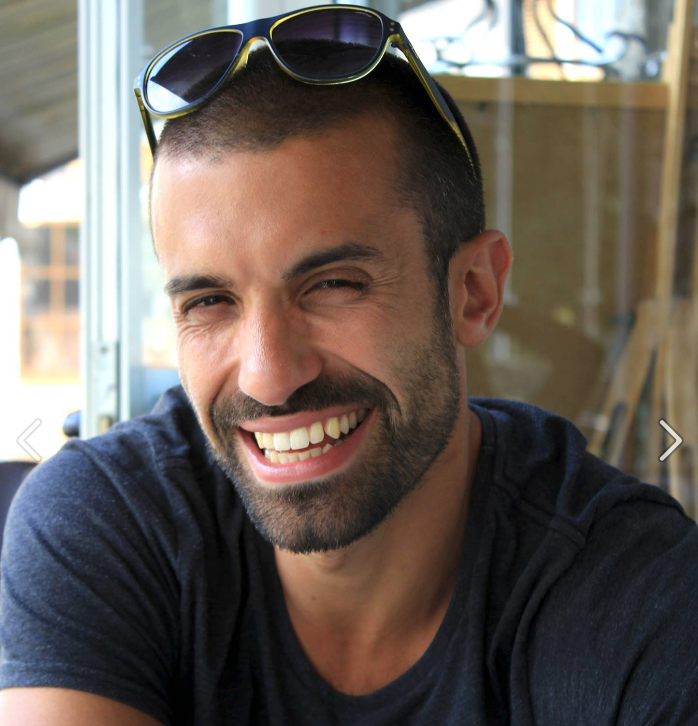}}]{Carlo Condo}
(M'15) received the M.Sc. degree in electrical and computer engineering from the Politecnico di Torino and The University of Illinois at Chicago, Chicago, IL, USA in 2010, and the Ph.D. degree in electronics and telecommunications engineering from the Politecnico di Torino and IMT Atlantique in 2014.
From 2015 to 2017, he was a Post-Doctoral Fellow with the ISIP Laboratory, McGill University, where he was a McGill University Delegate at the 3GPP meetings for the fifth-generation wireless systems standard (5G) in 2017. Since 2018, he has been a Researcher with the Communication Algorithms Design Team, Huawei Paris Research Center. His Ph.D. dissertation received a mention of merit as one of the five best of 2013/2014 by the GE association. He was a recipient of two conference best paper awards (SPACOMM 2013 and ISCAS 2016).
His research is focused on channel coding, design and implementation of encoder and decoder architectures, digital signal processing, and machine learning.
\end{IEEEbiography}

\begin{IEEEbiography}[{\includegraphics[width=1in,height=1.25in,clip,keepaspectratio]{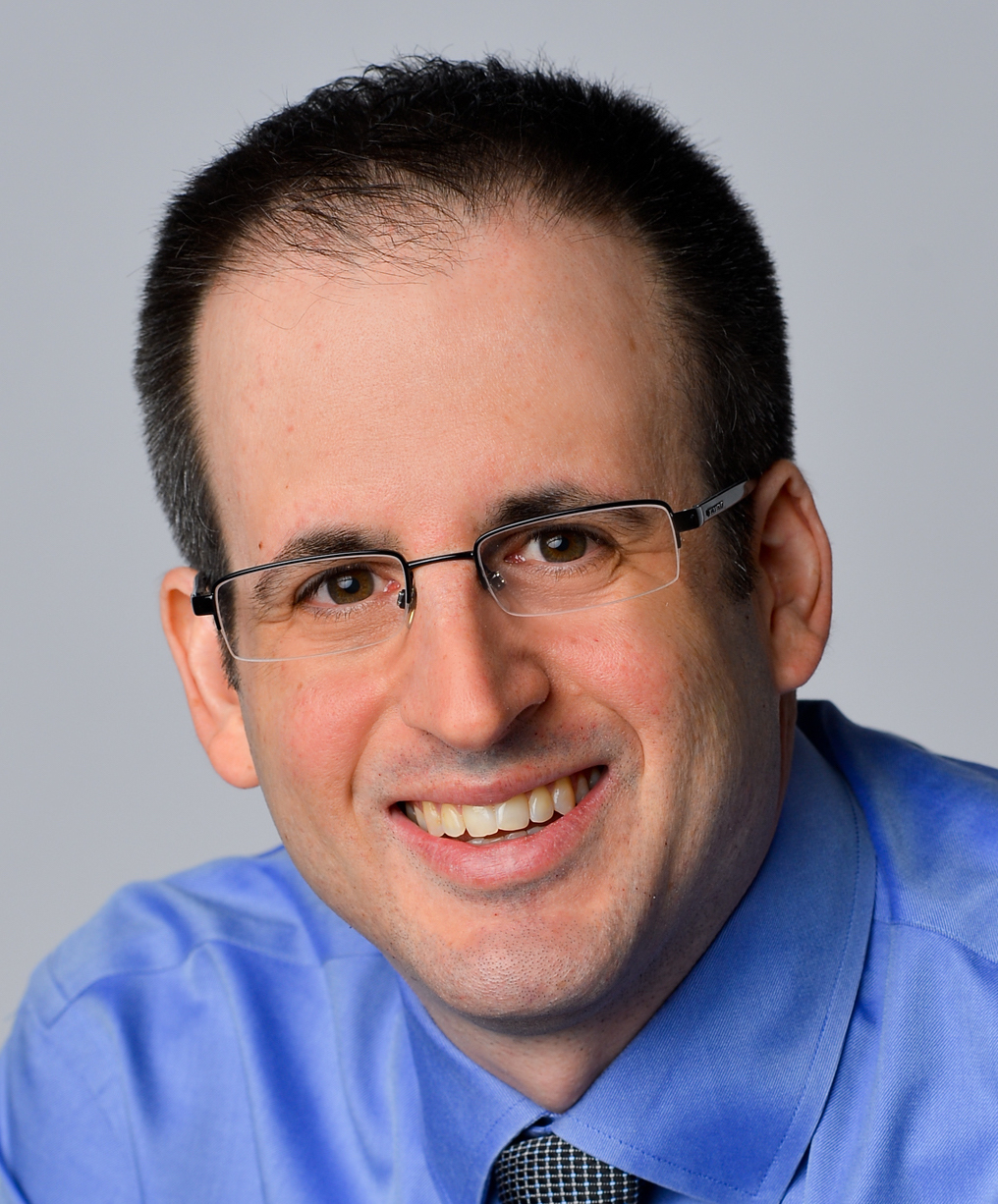}}]{Warren J. Gross}
(S'92--M'04--SM'10) received the B.A.Sc. degree in electrical engineering from the University of Waterloo, Waterloo, ON, Canada, in 1996, and the M.A.Sc. and Ph.D. degrees from the University of Toronto, Toronto, ON, Canada, in 1999 and 2003, respectively. He is currently a Professor and the Chair of the Department of Electrical and Computer Engineering, McGill University, Montreal, QC, Canada. His research interests are in the design and implementation of signal processing systems and custom computer architectures.

Dr. Gross served as the Chair for the IEEE Signal Processing Society Technical Committee on Design and Implementation of Signal Processing Systems. He served as the General Co-Chair for the IEEE GlobalSIP 2017 and the IEEE SiPS 2017 and the Technical Program Co-Chair for SiPS 2012. He also served as an Organizer for the Workshop on Polar Coding in Wireless Communications at WCNC 2017, the Symposium on Data Flow Algorithms and Architecture for Signal Processing Systems (GlobalSIP 2014), and the IEEE ICC 2012 Workshop on Emerging Data Storage Technologies. He served as an Associate Editor for the IEEE TRANSACTIONS ON SIGNAL PROCESSING and as a Senior Area Editor. He is a Licensed Professional Engineer in the Province of Ontario. 
\end{IEEEbiography}

\end{document}